\documentclass[smallextended]{svjour3}       

\usepackage{multirow}
\usepackage{balance}
\usepackage{graphicx}
\usepackage{alltt}
\usepackage{relsize}
\usepackage{booktabs}
\usepackage{array}
\usepackage{amsmath}
\usepackage{verbatim}
\usepackage{adjustbox}
\usepackage{afterpage}
\usepackage{lscape}

\usepackage[tight,footnotesize]{subfigure}

\usepackage{amsfonts}
\usepackage{amssymb}
\usepackage{boxedminipage}
\usepackage{float}
\usepackage{fancyvrb}
\usepackage[dvipsnames]{xcolor}
\usepackage[colorlinks=true]{hyperref}
\hypersetup{
linkcolor=blue
,citecolor=blue
,filecolor=blue
,urlcolor=blue
,menucolor=blue
,runcolor=blue
,linkbordercolor=blue
,citebordercolor=teal
,filebordercolor=blue
,urlbordercolor=blue
,menubordercolor=blue
,runbordercolor=blue
}
\usepackage{balance}
\usepackage{url}
\usepackage{fancybox}
\usepackage{listings}
\usepackage{array}
\usepackage{lscape}
\usepackage{textcomp}
\usepackage{makecell}
\usepackage{rotating}
\usepackage{pdflscape}

\definecolor{mygray}{rgb}{0.7,0.7,0.7}

\newcommand{\heng}[1]{\textcolor{red}{{\it [Heng: #1]}}}
\newcommand{\xingfang}[1]{\textcolor{teal}{{\it [Xingfang says: #1]}}}

\newcommand{\response}[2]{#2}

\newcommand{\accepted}[1]{}
\newcommand{\mresponse}[2]{#2}

\usepackage{framed}
\usepackage{dcolumn}
\usepackage{tikz}
\usepackage{tcolorbox}
\tcbuselibrary{skins, breakable, theorems}
\definecolor{fbtitle}{HTML}{636463}
\definecolor{fbbg}{HTML}{F2F2F2}

\newtcbtheorem{answer}{Answer}%
  {enhanced, breakable,
    colback = fbbg, colframe = gray, colbacktitle = fbtitle,
    attach boxed title to top left = {yshift = -2.2mm, xshift = 0mm},
    boxed title style = {rounded corners},
    separator sign none,
    fonttitle = \sffamily\bfseries}{qst}

\newtcbtheorem{summary}{Summary}%
  {enhanced, breakable,
    colback = fbbg, colframe = gray, colbacktitle = fbtitle,
    attach boxed title to top left = {yshift = -2.2mm, xshift = 0mm},
    boxed title style = {rounded corners},
    separator sign none,
    fonttitle = \sffamily\bfseries}{qst}

{\begin{list}{}%
         {\setlength{\leftmargin}{#1}}%
         \item[]%
}
{\end{list}}

{ \medskip
  \noindent
  \let\emph=\textbf
  \begin{boxedminipage}{\columnwidth}\em}%
{ \end{boxedminipage}}

\newdimen\qdx
\newdimen\qda
\newdimen\qdb
\def\rrrr#1#2#3#4{\newdimen\qd\qd=#4 
\qdx=\qd\multiply\qdx by 5\divide\qdx by 4
\qda=\qd
\qdb=\qd
\multiply\qda by #1\divide\qda by #3\multiply\qdb by #2\divide\qdb by #3\advance\qdb by -\qda
    \leavevmode\hbox to \qdx{\hfil\vbox{%
    \hbox{\vrule\vbox{\hrule\hbox to 1\qd
            {\vrule depth0pt height0.7ex width \qda\color{mygray}%
 \vrule depth0pt height0.7ex width \qdb\hfill}\hrule}\vrule}
    }\hfil}}

  \renewenvironment{thebibliography}[1]{%
    \begin{oldthebibliography}{#1}%
      \setlength{\parskip}{0ex}%
      \setlength{\itemsep}{0ex}%
  }%
  {%
    \end{oldthebibliography}%
  }

\usepackage{graphicx}
\usepackage{framed}
\usepackage{lscape}
\usepackage{setspace}
\usepackage{tabularx}
\usepackage{multirow}
\usepackage{stfloats}
\usepackage[normalem]{ulem}
\usepackage{url}
\usepackage[flushleft]{threeparttable}
\usepackage{comment}
\usepackage[round,sort]{natbib}

\usepackage{pgfplots}
\pgfplotsset{compat=1.15}
\usetikzlibrary{positioning}
\usepackage{vcell}

\makeatletter
\renewcommand\hyper@natlinkbreak[2]{#1}
\makeatother

\def\makeheadbox{{%
\hbox to0pt{\vbox{\baselineskip=10dd\hrule\hbox
to\hsize{\vrule\kern3pt\vbox{\kern3pt
\hbox{\bfseries Empirical Software Engineering}
\hbox{This is a pre-copyedit version of this article.}
\kern3pt}\hfil\kern3pt\vrule}\hrule}%
\hss}}}

\begin{document}

\title{Characterizing and Classifying Developer Forum Posts with their Intentions
}

\titlerunning{Characterizing and Classifying Developer Forum Posts with their Intentions}        

\author{Xingfang Wu         \and
        Eric Laufer \and
        Heng Li \and
        Foutse Khomh   \and 
        Santhosh Srinivasan \and
        Jayden Luo
}


\institute{Xingfang Wu, Heng Li, Foutse Khomh                 \at
              Department of Computer Engineering and Software Engineering \\
              Polytechnique Montréal \\
              Montréal, Québec, Canada \\
              \email{\{xingfang.wu, heng.li, foutse.khomh\}@polymtl.ca} \\\\
              Eric Laufer, Jayden Luo
\at
              Peritus.ai Canada Inc.
              Montréal, Québec, Canada \\
              \email{\{eric, jayden\}@peritus.ai} \\\\
              Santhosh Srinivasan
\at
              Peritus.ai, Inc.
              Palo Alto, California, United States \\
              \email{sms@peritus.ai}
}

\date{Received: date / Accepted: date}

\maketitle

\begin{abstract}
With the rapid growth of the developer community, the amount of posts on online technical forums has been growing rapidly, which poses difficulties for users to filter useful posts and find important 
information. 
Tags provide a concise feature dimension for users to locate their interested posts and for search engines to index the most relevant posts according to the queries. 
Most tags are only focused on the technical perspective (e.g., program language, platform, tool). In most cases, forum posts in online developer communities reveal the author’s intentions to solve a problem, ask for advice, share information, etc. The modeling of the intentions of posts can provide an extra dimension to the current tag taxonomy.
By referencing previous studies and learning from industrial perspectives, we create a refined taxonomy for the intentions of technical forum posts. Through manual labeling and analysis on a sampled post dataset extracted from online forums, we understand the relevance between the constitution of posts (code, error messages) and their intentions. Furthermore, inspired by our manual study, we design a pre-trained transformer-based model to automatically predict post intentions. 
The best variant of our intention prediction framework, which achieves a Micro F1-score of 0.589, Top 1-3 accuracy of 62.6\% to 87.8\%, and an average AUC of 0.787, 
outperforms the state-of-the-art baseline approach. 
Our characterization and automated classification of forum posts regarding their intentions may help forum maintainers or third-party tool developers improve the organization and retrieval of posts on technical forums.


\end{abstract}

\section{Introduction}\label{sec:intro}

Online technical communities such as Stack Overflow have been growing rapidly in recent years.
By Apr 2023, more than 24 million posts
and 35 million answers exist on Stack Overflow only\footnote{\label{so_amount}\url{https://stackexchange.com/sites?view=list}}, let alone a large number of posts on the innumerable private channels and dedicated forums that are not universal and general for all developers.
The rapid growth of online developer communities demands more efficient and advanced approaches to managing content and providing users with more precise query results and accurate recommendations. 

\response{R2.1}{
Tags often serve as a starting point for developers to investigate the topics discussed in forums~\citep{barua2014developers}. Tags function as crucial meta information, aiding in the categorization of content~\citep{maity2019deeptagrec}. This empowers users to efficiently filter out irrelevant posts, enabling them to swiftly refine their search. On the other hand, tags enable the recommendation of posts to specific user groups by aligning with their user portrait, which is constructed from their activity history (including browsing history, answered questions, etc.) associated with those specific tags, ensuring a more personalized content recommendation~\citep{huang2017expert, greco2018stackintheflow, guo2008tapping}. Efficient recommendations with tags have the potential to enhance the visibility of a question, increasing the likelihood of a swift response from domain experts~\citep{yazdaninia2021characterization}.
}

\response{R2.1}{
Most online technical communities provide users with pre-defined tags when posting. \response{R1.4}{However, most of these pre-defined tags prioritize technical aspects, including programming languages, platforms, and tools~\citep{chen2019modeling, beyer2020kind}. For example, tags related to programming languages, such as JavaScript, Python, Java, C\#, and PHP predominate among the most popular tags on Stack Overflow.~\footnote{https://stackoverflow.com/tags}} In some cases, users are allowed to input their customized tags. However, the quality of customized tags depends largely on users’ level of expertise. Customized tags may suffer from improper granularity and redundancy~\citep{maity2019deeptagrec, soapplyingtags}.
}

\response{R2.1}{
These tags can sometimes be too wide-ranging or specific, leading to poor post distinction~\citep{soapplyingtags}. Moreover, the presence of similar or duplicated tags complicates the process of recommending posts practically, making it more challenging to provide accurate and useful post recommendations~\citep{maity2019deeptagrec}. Therefore, tag recommendation approaches are researched and developed for online technical communities~\citep{al2010fuzzy, wang2015tagcombine,hong2017efficient, liu2018fasttagrec, zhou2019deep, li2020tagdc}. Most of these approaches are based on textual features of posts and adopt natural language processing to recommend potential post tags.
These tag-recommendation approaches, which only focus on technical topics of posts, have achieved good performances both on public datasets and real application scenarios~\citep{he2022ptm4tag}. 
}

However, a recent study~\citep{beyer2017analyzing} suggests that it may not be sufficient to only consider technical issues and topics when analyzing the questions proposed by developers. It stands a better chance of getting more insights into 
the posts if we investigate reasons why questions are asked~\citep{allamanis2013and}. These reasons can provide an extra dimension for developers to find solutions in online communities and support better recommendations of auxiliary tools for developers~\citep{beyer2020kind}. In this paper, we refer to these reasons as \textit{intentions}.
\response{R1.3}{To exemplify the distinctions between technique-oriented tag taxonomies and an intention-based taxonomy for technical posts, we present a concrete example. Suppose a novice programmer posts a question seeking a solution to address an error in their Python code related to data stored in an array data structure. Additionally, the programmer expresses a desire to find relevant tutorials to enhance their comprehension. The existing tag taxonomy would likely tag the question with `python' and `array' tags, emphasizing the technical aspects involved. On the other hand, the post embodies dual purposes: first, it serves as a request for assistance in resolving an \textit{error} within the Python language; second, the programmer seeks \textit{learning} resources about the related knowledge, particularly regarding data structures. Consequently, within an intention taxonomy, this post may be labelled with tags like `error' and `learning'. This distinction underscores the diverse nature of the questioners' intentions, extending beyond the limited technical categorization offered by the current tag taxonomy.}



Previous works have proposed different taxonomies of online technical posts with different focuses and approaches~\citep{allamanis2013and, beyer2020kind, treude2011programmers, beyer2014manual}. However, most of the previous studies suffer from several drawbacks; in most studies, only posts from a single technical community are considered, some of which only focus on a single specialized domain~\citep{beyer2014manual, beyer2020kind}.
As far as we know, most works are empirical studies carried out by researchers in academia. Therefore, a gap may exist between existing works and actual industrial practices. \response{R2.2}{In our study, our primary goal is to narrow the gap by integrating industry insights into the construction of an intention detection approach for technical forum posts. To achieve this, we've outlined four key objectives. Firstly, we aim to \textbf{understand the distribution and placement of post content}, providing crucial insights for constructing our post-analysis workflow. Secondly, our focus is to \textbf{identify latent intentions} of technical posts and \textbf{their correlations with content distribution}, aiding us in crafting an intention taxonomy and detection approach that fits practical scenarios. Our third objective revolves around \textbf{crafting an intention taxonomy} that incorporates suggestions and needs from the industry. Finally, our fourth objective entails \textbf{constructing an intention detection approach} based on our refined taxonomy. To achieve these objectives, we base our study on a dataset collected from an in-use recommendation system and continually integrate feedback from our industrial collaborator.}
We first conducted a qualitative study to understand the common posting practices in online technical communities. In the qualitative study, we examined a dataset which contains posts from different technical communities of different platforms. With the dataset, we manually look into the general composition of forum posts and the usage of different facilities (e.g., code block, image and etc.) supported by the platforms. 

To obtain a taxonomy that can better serve the industrial application scenarios and use cases, we reviewed previous works and their suggested taxonomies. We evaluated the significance (e.g., the number of relevant posts found, usefulness regarding to industrial use cases) of existing intention categories. Based on this evaluation, significant intention categories were identified, which were reused and adapted accordingly.
Our work is performed on a dataset of forum posts provided by our industrial partner that covers multiple developer communities (e.g., Stack Overflow, Discourse forums, etc.).

Furthermore, we manually annotate the intentions of posts following a rigorous process according to the resulting taxonomy of technical forum post intentions. 
Based on the findings and insights from the qualitative study, we propose an intention prediction framework for technical online posts. 
In the framework, we employ transformer-based pre-trained language models to generate embeddings for both title and description of posts. 
In addition to the textual descriptions of the posts, we also consider structural features such as the category of content contained in the code blocks.
To improve the performance, we further fine-tuned the pre-trained model with our annotated dataset. To examine the effectiveness and get a better understanding of the intention prediction framework, we proposed the following research questions (RQs):


\textbf{RQ1: Which pre-trained model (PTM) works best in our framework?}

\textbf{RQ2: Can our framework benefit from fine-tuning the PTMs? Compared with the baseline models, how effective is our intention detection framework?}

\textbf{RQ3: Can the content category of code blocks really help the detection of post intentions?}


The major contributions of this work are as follows:

\response{R2.3}{
    \begin{enumerate}
      \item Our results from the qualitative study provide insights into the composition of technical posts and the correlation between the composition and the intention of posts. These findings can provide guidance for future approaches on technical forum post analysis.
      \item We devise a post intention taxonomy by incorporating suggestions and needs from the industry. Based on the taxonomy, we construct and publish a technical post dataset with intention annotations, which future researchers and practitioners can utilize to build and evaluate their intention detection approaches.
      \item We propose an intention detection approach, leveraging and fine-tuning pre-trained language models, which outperformed the baselines. To ensure reproducibility, we have made the code publicly accessible.
      \item Our evaluation of six PTMs (including both general-purpose and domain-specific ones) in the task of intention detection provides guidance for choosing PTMs in processing technical forum post data. \mresponse{R2:}{Together with the findings from the qualitative study, we provide some recommendations for practitioners when developing and using technical forums.}
      \item We provide insights for technological implementations and endeavours in industry-academia collaborations, drawing from lessons learned during the construction of our intention detection approach and the co-construction process with our industrial partner.
    \end{enumerate}
}

The paper is structured as follows: Section~\ref{sec:pre} introduces the qualitative study on online technical posts and our proposed taxonomy of post intentions. Section~\ref{sec:method} specifies our approach to predicting post intention with textual and structural information of posts, and the details of our proposed framework and training process are detailed. In section~\ref{sec:eval}, we evaluate our proposed framework by three research questions (RQs) and analyze the results. \response{R1.10}{Section~\ref{sec:industrial} presents insights from collaborative industry endeavours, highlights key findings from our experiments, and provides suggestions for forum users regarding posting behaviors.} Section~\ref{sec:threats} identifies the threats to the validity of our study. Section~\ref{sec:related} surveys related works. At last, we conclude and summarize our work in section~\ref{sec:conclusions}.

\section{Qualitative Study on Technical Posts and Their Intention}
\accepted{\heng{not preliminary, qualitative?}}
\label{sec:pre}

\accepted{\heng{this paragraph does not belong to overview. maybe put it as a separate sub-section ``Background'' before the overview sub-section?}}
\subsection{Posts in online technical communities}

\begin{figure}[!t]
    \centering
	\includegraphics[scale=0.7]{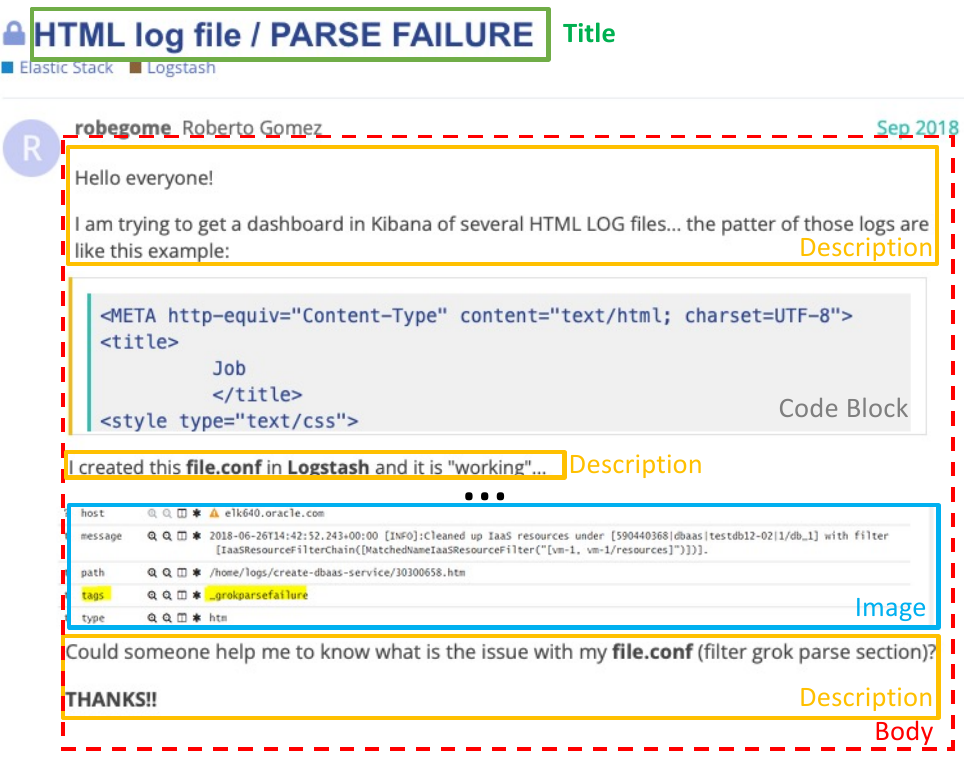}
	\caption[Discourse]{An example post from elastic.co, a Discourse-based online community.}
	\label{fig:exemple_df}
\end{figure}

To provide some background for our study, we first briefly introduce the common post structure in online technical communities. While our study is not constrained to any particular online community or technical domain, technical posts typically adhere to a comparable structure and composition with minor deviations resulting from variations in platforms. Fig.~\ref{fig:exemple_df} shows an example of posts on Discourse Forum. Generally, a post contains a short \textit{title} which presents the main purpose or topic of the post. There are usually some detailed \textit{descriptions} written in natural language in the \textit{body} part of the post. \textit{Code blocks}, \textit{images}, or even \textit{files} may be appended to a post as supplementary materials to provide a clearer picture for readers to understand the situation. However, not all of them are supported by the different platforms of online communities. These supplementary materials may contain various formats, varying by users with different usage habits. For example, code blocks are supposed to contain code snippets, while in reality, some users may not follow the norms and put some descriptions in natural language in code blocks. Last but not least, most online platforms support \textit{tags}, which helps organize the contents in the communities.


\subsection{Overview of the Manual Study}\label{subsec:man}

\begin{figure*}[!ht]
    \centering
	\includegraphics[width=\textwidth]{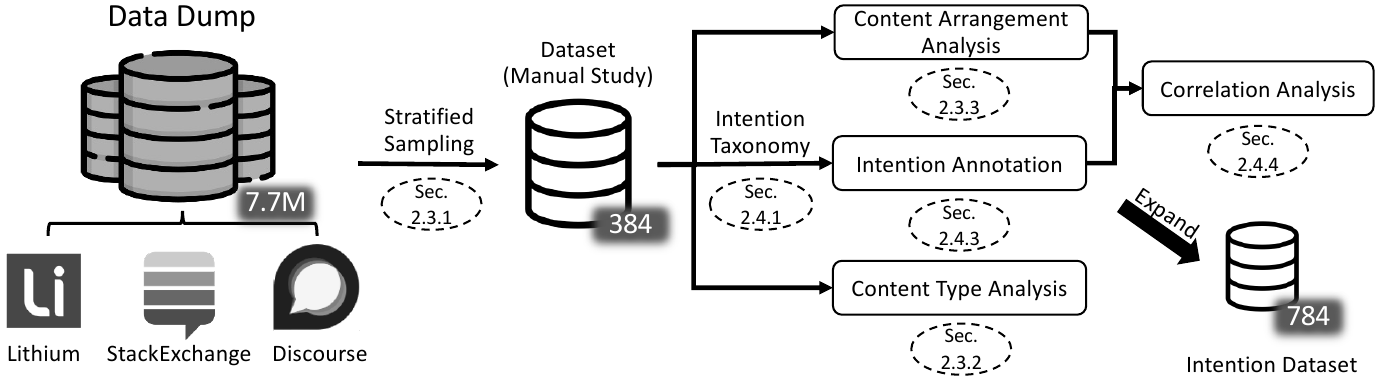}
	\caption[Manual Study]{\response{R2.5}{An overview of our manual study process.}}
	\label{fig:manual}
\end{figure*}


\response{R2.5}{Figure~\ref{fig:manual} shows an overview of our manual study, with each step aligned to a corresponding section for easy navigation. \response{R2.4}{Generally, the entire manual study belongs to the \textit{Sample Study~\citep{stol2018abc}}, encompassing the examination of content types and intentions within our sampled dataset. Initiated with a data dump from our industrial partner, our manual study begins by sampling this dataset to create a representative subset. Within this subset, we first conducted a qualitative manual study on the content and its arrangement of technical posts, and we further studied the intentions of technical forum posts from an industrial point of view. For the content and arrangement study, we first developed a content type taxonomy through an \textit{open coding} approach \citep{khandkar2009open} and annotated the sampled posts accordingly. Utilizing the outcomes from the content type analysis, we delved further into understanding how these content types were positioned within posts, employing an additional manual annotation process. In our manual study of post intention, we also employed the \textit{open coding} method. However, we scrutinized the sample dataset and consulted existing technical post taxonomies and industrial feedback to construct our intention taxonomy. Based on this taxonomy, we annotated the sample dataset, involving a systematic analysis and categorization of posts into intention categories. Finally, we explored correlations between the occurrence of content types and intentions, aiming to derive insights for designing intention detection approaches. We then expanded our annotated dataset to encompass 784 samples, essential for crafting our intention detection approach. In this section, we delve into each step of this process, detailing our methodology and findings.}}


\subsection{A Manual Study of Content Types and Their Arrangement in Technical Forum Posts}\label{subsec:structure-study}

\response{R1.6}{In this section, we focus on an examination of the various types of content present in forum posts, such as code snippets, error messages, and images, and how these elements are arranged. Our focus on how these elements are organized aims to provide a clearer understanding of how information is typically structured in these forum posts. Insights from this analysis have the potential to guide the development of automatic intention detection approaches for technical posts, potentially enhancing their ability to interpret and categorize posts more effectively and discern questioners' intentions in online technical communities.
}

\subsubsection{Dataset and Data Sampling}
\label{subsubsec:dataset}
We constructed our studied dataset by sampling from a data dump provided by our industrial partner. \response{R3.2}{The data dump was constructed during the time frame from June 25, 2020, to April 5, 2022. The dump contains primary posts (initial topic-setting posts) from different sources (i.e., online communities), mainly from three different platforms: Stack Exchange\footnote{\label{fnote:so} A question-and-answer website that covers a wide range of topics and domains. The data dump only contains contents from selected technical subforums.}, Lithium\footnote{\label{fnote:li} A forum software developed by Lithium Technologies.} forums and Discourse\footnote{\label{fnote:dc} An open-source forum software.} forums. The acquisition process of this data dump was facilitated by our partner's own web crawlers. Due to NDA, we could not share more details about the data crawling approaches. From around 7.7 million primary posts in the data dump, we adopted a stratified random sampling strategy to sample 384 posts in order to obtain precise estimates of the characteristics of posts from different platforms. The sampled amount is based on a 95\% confidence level and a 5\% margin of error~\citep{boslaugh2012statistics}. It's worth noting that in our sample size calculation, we considered only the question posts from the data source. We did not include answer posts in our calculations.}

However, during the manual annotation process, we found that 20 posts were no longer accessible online. As the dump does not contain some media (e.g., image), we could not conduct the content annotation for these posts. Thus,~\response{R1.2 R3.2}{we randomly sampled an additional 20 posts that were available online during this research. Finally, we constructed a dataset with 384 posts, of which 198 were from Stack Exchange, 137 from Lithium and 49 from Discourse forums, as shown in Table~\ref{tab:post-dist}.}~\response{R2.9}{As the annotation for content types and their arrangement did not involve significant discrepancies, and any disagreements arose from mistakes made by raters, which were subsequently corrected by a third rater, we did not measure the inter-rater agreements for these two annotation process.}

\begin{table}
\begin{minipage}[!t]{.49\columnwidth}
  \caption{Distribution of Sampled Posts by Data Source~\response{R1.2 R3.2}{Updated.}}
  \label{tab:post-dist}
  \centering
  \setlength{\tabcolsep}{0.5mm}{

    \begin{tabular}{ccc} 
    \toprule
    \textbf{Data source} & \textbf{Total} & \textbf{\textbf{Sampled}}  \\ 
    \hline
    Stack Exchange       & 4,058,490            & 198 (52\%)                       \\
    Lithium              & 2,694,643            & 137 (36\%)                       \\
    Discourse            & 948,580              & 49 (12\%)                        \\
    \bottomrule
    \end{tabular}
    }
\end{minipage}
\begin{minipage}[!t]{.42\columnwidth}
  \caption{Distribution of Content Types in the Post Dataset~\response{R1.2 R3.2}{Updated.}}
  \label{tab:content_dist}
  \centering
  \setlength{\tabcolsep}{0.6mm}{
  \begin{tabular}{l|c} 
\hline
\multicolumn{1}{c|}{\textbf{Content}} & \textbf{Percentage}  \\ 
\hline
Code                                      & 26.8\%               \\
Error message                             & 15.9\%               \\
Image                                     & 10.4\%                \\
Config                                    & 8.9\%                \\
Command line                              & 6.5\%                \\
Others                                    & 10.4\%                \\
\hline
\end{tabular}
}
  \end{minipage}
\end{table}

\response{R2.6}{}
\subsubsection{Manual Analysis of Content Types}
\label{subsubsec:contenttype}
\noindent \textit{\textbf{Introduction}}
The primary goal of this analysis is to gain insights into the utilization of various content types, such as code snippets and images, within technical forums.

\noindent \textit{\textbf{Methodology}}
The manual study of content types includes both annotation and result interpretation. Three authors (for example, A1, A2, and A3) participate in the annotation for content types of technical posts, involving the following three phases:

\begin{enumerate}
    \item A1 and A2 went over the instances in the sample dataset together. Through the initial inspection which involved an \textit{open coding} approach~\citep{khandkar2009open}, the two authors summarized a list of content types that posts contain. And then, the two authors annotated 100 random samples collaboratively to reach a consensus on annotation standards.
    \item A1 and A2 independently annotated sampled posts with the consensus reached by the discussion in Phase 1.
    \item After finishing the individual works, A3 checked and summarized the results.
\end{enumerate}

\noindent \textit{\textbf{Results}}
We count the occurrence of different types of content authors use to describe their issues. The content types mainly include natural language, code (both inline and multi-line code snippets), error message ( stack trace, log, error output), image, command line and others. These types of content can appear in different parts of a post. For example, logs can appear both in the description and code blocks. It is worth noting that we count the occurrence of content no matter where they are in the posts, and we only count once if one type of content appears several times in a post.

All the posts in our dataset contain natural language to describe their issues. For other content types, we list the percentages of posts in Table~\ref{tab:content_dist}. \accepted{\heng{make sure the term in the table is consistent}}
\response{R1.2 R3.2}{Code (both inline or multi-line) is the most common (26.8\%) supporting content authors employ to provide extra information. 15.9\% of posts contain Error messages (stack trace, log, error output) in their body. As authors often truncate long stack traces and log sequences to their posts, we cannot distinguish among these content types in detail in most cases. Therefore, we merge these as \textit{Error message} here. Around 10\% of posts contain images. Configs are usually given in markup formats (e.g., XML, JSON, etc.) in 8.9\% of the posts. Only 6.5\% of posts contain command lines.} There are also some contents that do not belong to the previous categories, or there is not enough clue for us to recognize their types. We assign them in the \textit{Others} class.

\begin{framed}
\noindent \textbf{Finding 1}: Code snippets are the most common supplementary content for programming-related posts. Besides code, program outputs (e.g., stack trace, log, etc.), configs, and command lines are utilized to provide additional information regarding the issues of posts.
\end{framed}

\subsubsection{Manual Analysis of Content Arrangement}
\label{subsubsec:contentarr}
\noindent \textit{\textbf{Introduction}}
Previously in Section~\ref{subsec:structure-study}, we examined the prevalent content types utilized by authors to articulate their programming issues. This analysis focuses on how the questioners arrange contents of different types in their posts (e.g., code is put in the code block or code is shown in a screenshot). 

\noindent \textit{\textbf{Methodology}}
Based on the annotations obtained in the previous analysis process, we then delve deeper into understanding how these content types are positioned within posts. Since the annotation process for content type arrangements doesn't require complex judgments, it is relatively straightforward. To maintain accuracy, both A1 and A2 independently annotated all samples. Following this, A3 reviewed the results to rectify any mistakes.

\noindent \textit{\textbf{Results}}
\response{R1.2 R3.2}{
From our annotation of the code snippets, we found that most posts arrange their code snippets well: 90.6\% of the posts which contain code snippets utilize the code block as the container. Among the misuses, most are the cases in which users did not use inline code elements to mark their short code snippets (e.g., variable name, function name). 33.3\% of the posts that contain inline codes did not mark them correctly. Besides, only one post in our dataset utilizes a screenshot to present its code snippet. 6.0\% of the posts contain a code snippet that is not in a code block. We also found that in some cases, authors of developer forum posts may use the code block as a blockquote, in which they tend to put words from other sources, outputs of the program and other texts in natural language.
}

\response{R1.2 R3.2}{
Among the stack traces, which usually extend over multi-lines, 55.0\% are arranged in code blocks, 25.0\% are shown by screenshots, and only 20.0\% are mixed with descriptions in natural language. However, error messages or fragments of outputs, which are shorter in length, are more often (65.7\%) mixed with the description. Around half of the configs and command lines are arranged in code blocks (51.6\% and 52.0\%, respectively).
}

\begin{framed}
\noindent \textbf{Finding 2}: Authors of programming-related posts treat code blocks as a container and use them differently. Code blocks may contain various types of content other than code snippets. Authors tend to arrange their stack trace, configs, command lines and other programming-related textual information in code blocks.
\end{framed}

\accepted{\heng{the table may be rotated (with 2 rows) to save space}}

\accepted{\heng{supplement may not be a good term here, maybe use ``components'' or ``content''?}}

\accepted{\heng{use ``Arrangement'' or ``Presentation''?}}

\subsection{Intentions of Technical Forum Posts}
\label{subsec:intention}

\response{R1.1}{Previous works propose different taxonomies for technical posts~\citep{treude2011programmers, allamanis2013and, beyer2014manual, rosen2016mobile, beyer2020kind}. These works have analyzed technical post categories and motivations from different angles, considering the particular technical fields and their corresponding contexts. Some of the categories resulted from these studies express or are closely related to the intended purposes of technical posts. By incorporating the use cases and suggestions from our industrial partner, we review and adapt existing categories proposed in previous works to our proposed taxonomy that focuses on the intention aspects of technical posts. In our taxonomy, we have seven intention categories, as follows. We assign a keyword to each of these intention categories. We will use these keywords to mention these intention categories hereinafter to enhance conciseness and clarity.}

\afterpage{%
    \clearpage
    \begin{landscape}
    
    \begin{table}
    \centering
    \renewcommand{\arraystretch}{1.2} 
    \resizebox{.79\linewidth}{!}{
    \begin{threeparttable}
    \caption{Intention Categories of Online Technical Forum Posts}
    \label{tab:intentions}
    \begin{tabular}{@{}p{0.09\linewidth}p{0.17\linewidth}p{0.58\linewidth}p{0.16\linewidth}@{}}
    \toprule
    \textbf{Intention} \newline \textbf{Keywords} & \textbf{Definition} & \textbf{Snippet Examples} & \textbf{Related Prior} \newline \textbf{Definitions} \\
    \midrule
    Discrepancy & Seeking explanations for software behavior discrepancies not explicitly related to errors & Any ideas what I am doing wrong here? \textsuperscript{1}
    
    I don't understand why I cannot ping internet clients. \textsuperscript{2}
    
    why won't my css or js apply in Firefox? \textsuperscript{3} & \citet{treude2011programmers}; \citet{allamanis2013and}; \citet{beyer2014manual}; \citet{beyer2020kind}\\
    Explicit Error & Seeking solutions for errors or exceptions & But I get error Unexpected null value. I can't handle it, have someone had similar problem? \textsuperscript{4}
    
    It gives me this exception. \textsuperscript{5}
    
    Does anyone know what this error means? \textsuperscript{6} & \citet{treude2011programmers}; \citet{beyer2014manual}; \citet{beyer2020kind}\\
    Review & Looking for improved solutions or guidance to make well-informed decisions & I’ve completed ... exercise. Any feedback would be greatly appreciated. \textsuperscript{7}
    
    Should I concatenate all certificates ... for ... directive in NGINX. \textsuperscript{8}
    
    Here is how my service account is configured: ... if I am using kubectl auth can-i incorrectly. \textsuperscript{9} & \citet{treude2011programmers}; \citet{beyer2014manual}; \citet{beyer2020kind};\\
    Conceptual & Seeking information or explanations without concrete implementations & What are BigQuery audit logs supposed to produce? \textsuperscript{10}
    
    Is Terraform the official Infrastructure as code solution for IBM Cloud? \textsuperscript{11}
    
    What is the gRPC++ equivalent of the Go context.Background()? \textsuperscript{12} & \citet{beyer2014manual}; \citet{beyer2020kind}\\
    Learning & Seeking learning resources for libraries, tools, or programming languages & If I could get a detailed guide or a link to an existing one that would be amazing. \textsuperscript{13}
    
    I'm reading Vulkan Tutorial ... the "Subpass dependencies" section confused me a lot. \textsuperscript{14}
    
    I can’t seem to find any current documentation that discusses this. \textsuperscript{15} & \citet{allamanis2013and}; \citet{beyer2017analyzing}; \citet{beyer2020kind}\\
    How-to & Requesting step-by-step instructions for specific tasks & How to read utf16 text file to string in golang? \textsuperscript{16}
    
    Workflow to clean badly scanned sheet music. \textsuperscript{17}
    
    What do I need to do to make sure each group has its own directory ... \textsuperscript{18} & \citet{treude2011programmers}; \citet{allamanis2013and}; \citet{beyer2014manual}\\
    \bottomrule
    \end{tabular}
    
    \begin{tablenotes}[para,flushleft]
\item[1] StackOverflow (ID: \href{https://stackoverflow.com/questions/68442411}{68442411})
\item[2] Aruba Networks Community (ID: \href{https://community.arubanetworks.com/t5/Cloud-Managed-Networks/Aruba-Central-offline-devices-after-creating-DHCP-Server/td-p/438751}{438751})
\item[3] StackOverflow (ID: \href{https://stackoverflow.com/questions/31944197}{31944197})
\item[4] StackOverflow (ID: \href{https://stackoverflow.com/questions/67894563}{67894563})
\item[5] StackOverflow (ID: \href{https://stackoverflow.com/questions/65468209}{65468209})
\item[6] Cisco Community (ID: \href{https://community.cisco.com/t5/network-security/importing-error-in-mc-for-pix-in-vms/td-p/202789}{202789})
\item[7] Mozilla Discourse (ID: \href{https://discourse.mozilla.org/t/structuring-page-of-content-please-assess/96262}{96262})
\item[8] Server Fault (ID: \href{https://serverfault.com/questions/775101}{775101})
\item[9] Server Fault (ID: \href{https://serverfault.com/questions/1019782}{1019782})
\item[10] StackOverflow (ID: \href{https://stackoverflow.com/questions/34302214}{34302214})
\item[11] StackOverflow (ID: \href{https://stackoverflow.com/questions/56739646}{56739646})
\item[12] StackOverflow (ID: \href{https://stackoverflow.com/questions/61408251}{61408251})
\item[13] Server Fault (ID: \href{https://serverfault.com/questions/971124}{971124})
\item[14] StackOverflow (ID: \href{https://stackoverflow.com/questions/68004511}{68004511})
\item[15] Rancher Forums (ID: \href{https://forums.rancher.com/t/setup-guide-for-rancher-efs/5085}{5085})
\item[16] StackOverflow (ID: \href{https://stackoverflow.com/questions/15783830}{15783830})
\item[17] StackOverflow (ID: \href{https://stackoverflow.com/questions/68401240}{68401240})
\item[18] Cisco Community (ID: \href{https://community.cisco.com/t5/ip-telephony-and-phones/seperate-directories-same-ucm/td-p/1967272}{1967272})
\end{tablenotes}
    \end{threeparttable}
    }
    \end{table}
    
    \end{landscape}
}

\response{R1.1}{\subsubsection{Intention Taxonomy}} \label{subsubsec:texonomy}
\response{R1.1}{
\noindent \textit{\textbf{Intention 1 (Discrepancy) \mresponse{R1.1}{Seeking explanations for software behavior discrepancies that are not explicitly related to errors.}}} \label{para:disc}
The posts of this category contain questions about problems and unexpected behaviors of systems, services or code snippets which the questioner has no clue how to solve. The problems or unexpected behaviors are not necessarily associated with errors or exceptions and could instead be related to user errors. In previous works, this type of posts are categorized as~\textit{Do not work}~\citep{allamanis2013and} or~\textit{What is the Problem…?}~\citep{beyer2014manual}. In works~\citep{beyer2020kind, treude2011programmers}, their taxonomies also have this category.
}

\response{R1.1}{
\noindent \textit{\textbf{Intention 2 (Explicit Error) Seeking solutions for explicit errors or exceptions.}}
This category addresses problems related to exceptions or errors. Often, error messages, exceptions, and stack traces are attached to posts, and the questioners usually ask for help in finding the root cause of an exception and the solutions to fix an error. This category differs from the~\textit{Discrepancy} category by primarily focusing on troubleshooting and resolving specific errors or exceptions encountered in software. Unlike the~\textit{Discrepancy} category, which deals with a broader range of unexpected behaviors, this category is specifically tailored to address issues that are directly manifested by errors or exceptions. Questioners in this category seek assistance in pinpointing the root causes of these errors and soliciting effective solutions for rectifying them. The inclusion of error-specific information and the nature of the inquiries set this category apart as a specialized resource for those encountering error-related challenges in developing or using software. It is a common category shared by many previous works~\citep{beyer2014manual, treude2011programmers, beyer2020kind}.
}

\response{R1.1}{
\noindent \textit{\textbf{Intention 3 (Review) Looking for improved solutions or guidance to make well-informed decisions.}}
Typically, the questioners who ask questions in this category already have solutions for their problems. Their intentions are to validate their proposed decisions or to search for a better solution for accomplishing a task. Usually, authors will post their code snippet decisions for readers to review. Related categories proposed by previous works are~\textit{Decision Help}~\citep{treude2011programmers},~\textit{Better Solution}~\citep{beyer2014manual}, etc. This category also exists in \citet{beyer2020kind}.
}

\response{R1.1}{
\noindent \textit{\textbf{Intention 4 (Conceptual) Seeking background information, explanations, or a better understanding of subjects or technology aspects without concrete implementations.}}
This category of posts usually contains questions about abstract or non-implementation level concepts, such as design patterns, background information, or limitations about some libraries or devices. In some cases, the authors want to know whether it is feasible to do something with tools, libraries, or other supplements mentioned (i.e., limitations of tools, libraries, etc.). In previous work~\citep{beyer2014manual}, this category is mentioned as~\textit{Is it possible…?}. This category also exists in \citet{beyer2020kind}.}

\response{R1.1}{
\noindent \textit{\textbf{Intention 5 (Learning) Seeking learning resources.}}
This category usually features requests for documentation or tutorials on a specific library, tool, or programming language. Compared with~\textit{How-to} posts, posts in this category usually do not focus on a specific question and ask for solutions or instructions. Instead, they are seeking for support to learn on their own. This category is also proposed in~\citet{beyer2020kind}, and is the combination of \textit{Learning a Language/Technology}~\citep{allamanis2013and} and \textit{Tutorials/Documentation}~\citep{beyer2017analyzing}.
}

\response{R1.1}{
\noindent \textit{\textbf{Intention 6 (How-to) Requesting specific, step-by-step instructions for particular tasks.}}
This post category mainly asks for concrete instructions for a specific application scenario or a particular task to fulfill. This category subsumes post type \textit{API usage} or \textit{Interaction of API classes} proposed in~\citet{beyer2020kind} and~\citet{beyer2017analyzing}, respectively. Other works~\citep{treude2011programmers, allamanis2013and, beyer2014manual} also have similar or equivalent types for this category of posts.
}

\response{R1.1}{
\noindent \textit{\textbf{Intention 7 (Other) Other intentions.}}
We noticed some technical forum posts that did not fit into common categories. However, creating specific categories for them may be unproductive and could undermine recommendation systems’ effectiveness\mresponse{R1.1-3}{, based on the feedback from our industrial partner, as the number of these posts is not significant}. We group extra categories under \textit{Other} in our work. Table~\ref{tab:other_intention} presents a list of example intentions and example post titles that belong to the \textit{Other} category.
}

\begin{table}
\centering
\caption{\response{R1.1}{Examples of posts belonging to the \textit{Other} class}}
\label{tab:other_intention}
\resizebox{\linewidth}{!}{
\begin{threeparttable}
\begin{tabular}{|c|p{7cm}|l|}
\hline
Intention                                                                       & \multicolumn{1}{c|}{Definition}                                                                                           & \multicolumn{1}{c|}{Post Snippet Examples}                                                                                                                                  \\ 
\hline
\vcell{\begin{tabular}[b]{@{}c@{}}Requesting \\software resources\end{tabular}} & \vcell{Seeking access to resources for immediate use or application.}                                                     & \vcell{\begin{tabular}[b]{@{}l@{}}Are there any plugins/tools available to …?~\textsuperscript{1}\\Where can I download gcc …?~\textsuperscript{2}\end{tabular}}                                               \\[-\rowheight]
\printcellmiddle                                                                & \printcelltop                                                                                                             & \printcelltop                                                                                                                                                          \\ 
\hline
\vcell{Announcing}                                                              & \vcell{Informing or sharing news, updates, or events without seeking deeper understanding or background information.}     & \vcell{\begin{tabular}[b]{@{}l@{}}Release v0.46.0 New Features~\textsuperscript{3}\\We released new iOS versions for ... ~\textsuperscript{4}\\Read all about this latest release in this blog ...~\textsuperscript{5}\end{tabular}}                                                                                                                                   \\[-\rowheight]
\printcellmiddle                                                                & \printcelltop                                                                                                             & \printcelltop                                                                                                                                                          \\ 
\hline
\vcell{Discussing a topic}                                                      & \vcell{Open-ended queries or topics aimed at sparking conversation, sharing opinions, or seeking input from a community.} & \vcell{\begin{tabular}[b]{@{}l@{}}Are there any plans to increase this?~\textsuperscript{6}\\WPA2 Vulnerability Discussion~\textsuperscript{7}\\Anyone know if it is worth upgrading to 4.1.3b?~\textsuperscript{8}\end{tabular}}  \\[-\rowheight]
\printcellmiddle                                                                & \printcelltop                                                                                                             & \printcelltop                                                                                                                                                          \\ 
\hline
\vcell{\begin{tabular}[b]{@{}c@{}}Reporting a \\problem or a bug\tnote{*}\end{tabular}}                                                         & \vcell{Identifying, describing, and potentially addressing issues or bugs within software.}                               & \vcell{\begin{tabular}[b]{@{}l@{}}The word License is misspelled.~\textsuperscript{9}\\Settings place edit screen has a misspelled hint.~\textsuperscript{10}\end{tabular}}                                                                                                                                \\[-\rowheight]
\printcellmiddle                                                                & \printcelltop                                                                                                             & \printcelltop                                                                                                                                                          \\ 
\hline
...                                                                             & \multicolumn{1}{c|}{...}                                                                                                  & \multicolumn{1}{c|}{...}                                                                                                                                               \\
\hline
\end{tabular}
\begin{tablenotes}[para,flushleft]
\item[1] StackOverflow (ID: \href{https://stackoverflow.com/questions/248589}{248589})
\item[2] HPE Community (ID: \href{https://community.hpe.com/t5/system-administration/gcc-where-are-you/td-p/2787835}{2787835})
\item[3] Rancher Community (ID: \href{https://forums.rancher.com/t/rancher-release-v0-46-0/995}{995})
\item[4] Cisco Community (ID: \href{https://community.cisco.com/t5/cisco-support-via-mobile/cisco-technical-support-v3-8-new-features/td-p/2571411}{2571411})
\item[5] HashiCorp Discuss (ID: \href{https://discuss.hashicorp.com/t/boundary-0-2-0-is-live/23223}{23223})
\item[6] Paloalto Networks (ID: \href{https://live.paloaltonetworks.com/t5/general-topics/pa-4020-max-nat-rule-limit/td-p/39013}{39013})
\item[7] Aruba Networks Community (ID: \href{https://community.arubanetworks.com/t5/Wireless-Access/WPA2-Vulnerability-Discussion/td-p/310066}{310066})
\item[8] Cisco Community (ID: \href{https://community.cisco.com/t5/application-networking/waas-4-1-3b/td-p/1261498}{1261498})
\item[9] Cisco Community (ID: \href{https://community.cisco.com/t5/webex-administration/license-agreement-in-webex-teams-panics-american-workers-causes/td-p/4154656}{4154656})
\item[10] Roblox Developer Community (ID: \href{https://devforum.roblox.com/t/game-settings-place-edit-screen-has-a-misspelled-hint/705907}{705907})
\\
\item[*] This category differs from \textit{Explicit Error} and \textit{Descrepancy} in its primary focus on reporting issues or bugs for attention, rather than seeking immediate solutions for specific errors or unexpected behaviors.
\end{tablenotes}
\end{threeparttable}
}
\end{table}

\subsubsection{Manual Study of Post Intention}\label{subsubsec:postintention}

To further investigate the intentions behind technical QA posts and understand the correlation between post structure and intention, we conducted a manual study of post intention. This manual study process involves both annotation and result interpretation. Three authors (for example, A1, A2, and A3, including an expert from our industrial partner) participate in annotating the intentions for technical posts. Importantly, each intention is not exclusive, meaning one post can contain more than one intention. For instance, authors may request a solution to an error (i.e., \textit{Explicit Error}) while simultaneously seeking help to understand a related abstract concept (i.e., \textit{Conceptual}). Therefore, in the manual annotation process, we assigned multiple labels for posts with more than one intention. Similar to the manual analysis process described in Section~\ref{subsubsec:contenttype}, we adopted an open coding approach. We began by examining a set of sampled posts to identify recurring intentions. For example, we noticed a recurring theme of posts seeking help to find tutorials or other learning materials, which we labeled as \textit{Learning} and added to our initial intention categories. This iterative process enabled us to refine and categorize intentions progressively based on the content and context of the posts. The annotation typically involves the following three phases:

\begin{enumerate}
    \item The authors summarized a list of intentions and collaboratively annotated 100 random samples to establish an annotation consensus.
    \item A1 and A2 independently annotated sampled posts with the consensus reached by the discussion in Phase 1.
    \item After finishing the individual works, A1 and A2 compared the results, inter-rater agreements were measured, and any disagreement regarding the annotation was discussed to reach agreements. If the two authors could not reach a consensus, A3 got involved in the discussion, and the three authors voted and made the final decisions.
\end{enumerate}

Besides the sampled posts, we further annotated more posts in our data dump to acquire more training and test data for our proposed framework for automatically detecting post intentions (in Section~\ref{sec:method}). At last, we were able to extend our intention annotation dataset to the size of 784.

\begin{table}
\centering
\caption{Results of Intention Annotation}
\label{tab:intention_lb}
\begin{tabular}{cc} 
\hline
Intention   & \multicolumn{1}{l}{Number of post}  \\ 
\hline
Discrepancy & 149                                 \\
Explicit Error      & 150                                 \\
Review      & 86                                  \\
Conceptual  & 159                                 \\
Learning    & 23                                  \\
How-to      & 273                                 \\
Other       & 86                                  \\
\hline
\end{tabular}
\end{table}

\response{R2.9}{\subsubsection{Inter-rater Agreement}\label{subsubsec:intrat}
We measured the inter-rater agreement between two coders using Krippendorff's Alpha score~\citep{krippendorff2011computing} for the outcomes of Phase 2. Krippendorff's Alpha, a standard and flexible coefficient for measuring inter-coder agreement, takes the form of:
\begin{equation}
\alpha = 1 - \frac{D_o}{D_e}
\end{equation}
where $D_o$ represents the observed disagreement among coders, while $D_e$ signifies the disagreement anticipated by chance.
The scale, ranging from -1 to 1, signifies the level of agreement among raters, where -1 indicates perfect disagreement, 0 implies no agreement beyond chance, and 1 denotes perfect agreement. Since our annotation involves multiple intention classes, we decomposed a multi-category observation as multiple binary observations
and conducted the analysis as if it were a binary scenario. Following Phases 1 and 2, the Krippendorff's Alpha coefficients for all intention categories range between 0.62 and 0.81, indicating moderate to good agreements. All three authors achieved agreement on every sample following Phase 3. Hence, our manual labeling for intentions can be considered trustworthy.
}

\subsubsection{Results of Intention Annotation}
\begin{figure}[!ht]
    \centering
	\includegraphics[scale=0.50]{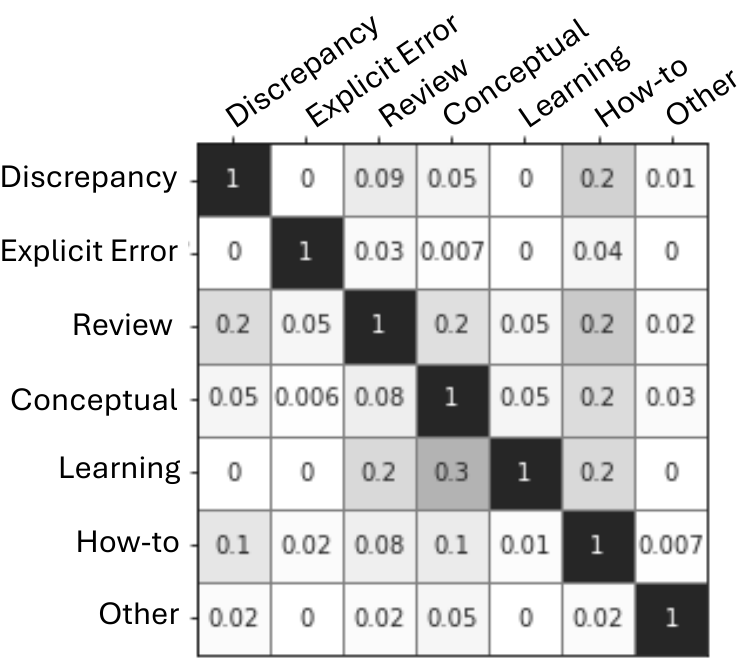}
    \vspace{2mm}
	\caption[Co-occurrence]{The co-occurrence matrix of intentions. Each row is divided by the number of posts of the corresponding intention. \accepted{\heng{the numbers along the diagonal should be 1, not 0?}}}
	\label{fig:int_co}
\end{figure}

The numbers of posts that belong to each intention class are shown in Table~\ref{tab:intention_lb}. We assigned 83\% of the posts with one label, 16\% with two labels, and only 1\% with three labels. \textit{How-to} is the most common intention, which accounts for 34.8\% of the posts. \textit{Review} and \textit{Learning} are the two least frequent intentions, which only occur in 11.0\% and 2.9\% of the posts, respectively. Moreover, we further counted the co-occurrence of the types of intentions when posts contain more than one type of intention. The co-occurrence matrix is shown in Figure~\ref{fig:int_co}. As the number of posts is unevenly distributed in the seven types, we divide each row\accepted{\heng{row?}} of the original co-occurrence matrix by the number of posts with the intention corresponding to that row. Thus, an element in row $i$, column $j$ shows the percentage of posts with intention $i$ that also have intention $j$. By summing up the elements in each row of the matrix, we can find that 69.6\% of the \textit{Learning} posts and 67.4\% of the \textit{Review} posts have other intentions. \textit{Learning} posts are usually also \textit{Conceptual}, \textit{Review} or \textit{How-to} posts. \textit{Review} posts are likely to be \textit{Discrepancy}, \textit{Conceptual} or \textit{How-to} posts. These co-occurrences are natural and reasonable. For example, when developers encounter an unexpected behavior of a program (\textit{Discrepancy}), they may provide their code snippets or operations for readers to check (\textit{Review}).

\accepted{\heng{not clear what is ``common cases''}}

\vspace{2mm}
\accepted{\heng{use italic font for all intentions?}}

\begin{framed}
\noindent \textbf{Finding 3}: Posts may have more than one type of intention.
\textit{How-to} is the most common type of intention, while the number of posts is unevenly distributed in the seven types of intentions. 
\end{framed}

\begin{table}[!ht]
\centering
\caption{Distribution of content types by intention types. \accepted{\heng{should make sure the columns match exactly the items in Table~\ref{tab:content_dist}}}}
\vspace{2mm}
\label{tab:intention_contents}
\resizebox{0.8\linewidth}{!}{
\begin{tabular}{l|c|c|c|c|c} 
\hline
\multicolumn{1}{c|}{\multirow{2}{*}{Intention}} & \multirow{2}{*}{Code} & \multicolumn{2}{c|}{Error Message}                                                                         & \multirow{2}{*}{Config} & \multirow{2}{*}{\begin{tabular}[c]{@{}c@{}}Command\\line\end{tabular}}  \\ 
\cline{3-4}
\multicolumn{1}{c|}{}                           &                       & \begin{tabular}[c]{@{}c@{}}Error\\text\end{tabular} & \begin{tabular}[c]{@{}c@{}}Stack\\trace\end{tabular} &                         &                                                                         \\ 
\hline
Discrepancy                                     & 43.0\%                & 9.3\%                                               & 1.2\%                                                & \textbf{15.1}\%         & \textbf{9.3\%}                                                          \\
Explicit Error                                          & 50.0\%                & \textbf{46.9}\%                                     & \textbf{26.6}\%                                      & 10.9\%                  & 7.8\%                                                                   \\
Review                                          & \textbf{65.8}\%       & 2.6\%                                               & 0.0\%                                                & 7.9\%                   & 5.2\%                                                                   \\
Conceptual                                      & 33.3\%                & 1.2\%                                               & 1.2\%                                                & 3.7\%                   & 6.2\%                                                                   \\
Learning                                        & 11.1\%                & 0.0\%                                               & 0.0\%                                                & 0.0\%                   & 0.0\%                                                                   \\
How-to                                          & 33.3\%                & 2.6\%                                               & 0.9\%                                                & 6.8\%                   & 5.1\%                                                                   \\
Other                                           & 4.0\%                 & 3.8\%                                               & 0.0\%                                                & 0.0\%                   & 3.8\%                                                                   \\
\hline
\end{tabular}
}
\end{table}

\subsubsection{Correlations Between the Occurrence of Certain Content Types and Post Intentions}\label{subsubsec:corr} From Table~\ref{tab:intention_contents}, we can find differences in the distributions of supplements among posts with different intention types. 65.8\% of Review posts and 50\% of Explicit Error posts have posted codes regarding their issues. The ratio is significantly lower for posts with other types of intention. As the nature of Explicit Error intention, posts of this type are more likely to have error texts or stack traces as their supplements. 46.9\% and 26.6\% of this type of posts contain error texts and stack traces separately. Also, we found that authors of Discrepancy posts are more likely to post their configurations for readers to address their issues.

\begin{framed}
\noindent \textbf{Finding 4}: There exists a correlation between posts’ intention types and their supplementary resources, which may serve as a feature for detecting the intentions of posts. The existence of different types of content may serve as a feature for detecting the intentions of posts.
\end{framed}




\newpage
\begin{summary*}{}{}
In addition to natural language descriptions, technical posts often contain different types of supplementary content (e.g., code, stack trace, etc.). Authors tend to arrange all these in code blocks. One technical post may have multiple intentions.
\accepted{\heng{need to briefly summarize intentions too.}}We observed a correlation between the presence of specific content types and the intentions.
\end{summary*}

\section{Automatically Detecting Post Intentions} \label{sec:method}

Inspired by the findings from our manual study and previous works, we propose a framework to detect the intentions for technical QA forum posts automatically. In this section, we describe in detail the proposed framework, which is based on a transformer-based PTM and formulates the process of detecting post intentions as a multi-class multi-label classification problem.

\subsection{Overview of the Framework}

\begin{figure*}[!t]
    \centering
	\includegraphics[scale=0.52]{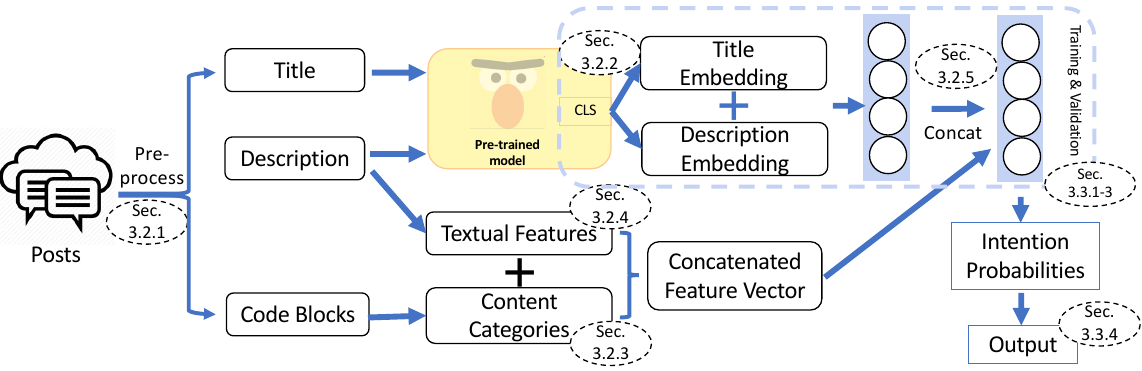}
	\caption[Framework]{An overview of our intention detection framework. The section numbers in the dashed circles correspond to the respective descriptions.}
	\label{fig:framework}
\end{figure*}

The overall structure of our proposed intention detection framework is illustrated in Figure~\ref{fig:framework}. Generally, the framework contains three processing stages: data pre-processing, feature extraction, and classification. During the pre-processing stage, we remove unexpected tokens (e.g., HTML tag) from the raw forum data. Then, in the feature extraction stage, we use a PTM as an encoder to generate embeddings\accepted{\heng{extract vector representations of?}} for the title and description of posts. The two embeddings are merged by a fully connected layer, the output of which is concatenated with a feature vector. The feature vector contains two parts of features: the content feature of code blocks and the textual features of the description of posts. The content feature is generated with a code block classifier, and textual features are generated with different metrics. Finally, the concatenated features are fed into a fully connected layer which outputs the classification results.


\subsection{Data Pre-processing \& Feature Extraction}

\subsubsection{Pre-processing} \label{subsubsec:prep}
According to the design of the online communities, posts may contain different HTML tags or other platform-specific tokens for the front-end formatting and presentation of the content. Code blocks are usually embedded in the \textit{Body} of the posts with specific tags (i.e., $<$pre$><$code$>$...$<$/code$><$/pre$>$ in Stack Overflow posts). In the pre-processing stage, we extract the content of code blocks and remove platform-specific tokens in the \textit{Body} of posts, which can be noise to the input of the PTM. \response{R3.6}{Typical pre-processing methods such as eliminating stopwords, performing stemming, and lemmatization are frequently utilized in natural language processing but are not mandatory for contextual embedding techniques. Stop words and declensions can sometimes provide contextual information for the model to better present the semantic information of texts, removing them may lead to a loss of information. Transformer-based PTMs can effectively manage variations in word forms and map them to continuous vector representations. A previous study has demonstrated that applying stopword removal has no effect on performance in their task~\cite{qiao2019understanding}. Therefore, we exclude these preprocessing strategies in our workflow as we adopt PTMs in our model.
}

\subsubsection{Generating embeddings with pre-trained models}

Bidirectional Encoder Representations from Transformers (BERT)~\citep{devlin2018bert} is a transformer-based architecture that is capable of capturing long dependency in natural language, and various transformer-based PTMs have been achieving state-of-the-art results in different natural language tasks~\citep{jin2020bert}. Besides, different PTMs which inherit the BERT architecture are developed and trained with program-related data to fulfill the tasks in software engineering and achieve promising results (e.g., CodeBERT~\citep{feng2020codebert}, BERTOverflow~\citep{tabassum2020code}). In our proposed intention detection framework, we employ PTMs released in the Hugging Face~\citep{wolf2019huggingface} to generate contextual embeddings for natural language content in QA posts.

\begin{figure}[ht]
    \centering
	\includegraphics[width=\textwidth]{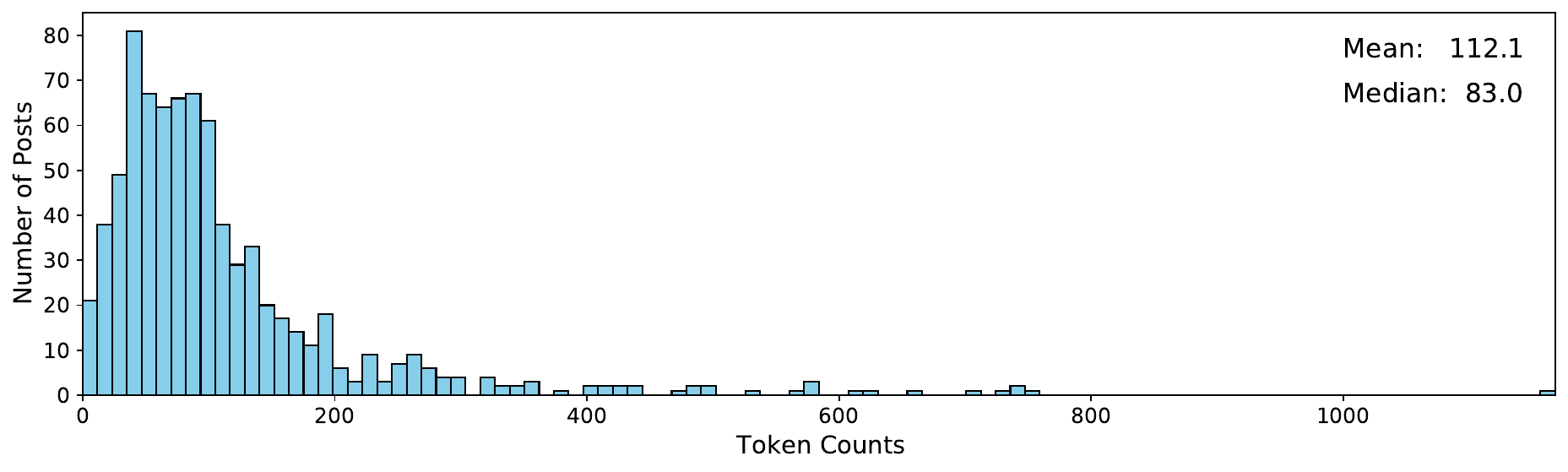}
	\caption[Distribution]{The distributions of description lengths of posts in our dataset.}
	\label{fig:dist_len}
\end{figure}

\noindent \textbf{Maximum Input Length} Due to the significant degradation in the performance of the BERT model in terms of the speed and accuracy of representing long documents, the authors of BERT set a limit to the input length of 512 sub-tokens~\citep{devlin2018bert}.
Sequences longer than the limit should be truncated. Choosing a proper maximum input length suitable for the data is essential for the framework’s performance in terms of speed and accuracy. 
In our framework, we feed the title and description separately to the tokenizer, followed by a PTM. All online communities have their limits for the post title, thus titles are of limited length. However, the length may vary in the description of posts. We found that most posts have a description part of fewer than 200 tokens. The average, median and maximum lengths are 112, 83 and 1168, respectively, in our dataset. Therefore, we set the maximum input sequence length as 256. For sequences longer than 256 tokens, we adopt the head-only truncation.
The description refers to the preprocessed \textit{Body} of posts, in which unexpected tokens and code blocks are removed. Therefore, we census the description length in our sampled dataset. 
The distributions of description lengths are shown in figure~\ref{fig:dist_len}.


\noindent \textbf{Pooling Strategy} 
After feeding the tokenized title and description to the PTM, word embeddings are generated for tokens. To acquire embeddings to represent the title and the description, we need to aggregate the embeddings with a pooling strategy. The common ways for that include: (1) using the output of the first \textit{$<$CLS$>$} token, (2) applying average or max pooling across each dimension of last hidden state embeddings~\citep{reimers2019sentence}, (3) applying pooling on a concatenation of last few layers~\citep{devlin2018bert}.
However, there is no clear guideline on which pooling strategy should be used in all application scenarios and all PTMs~\citep{reimers2019sentence, devlin2018bert}. 
We choose to use the output of the first \textit{$<$CLS$>$} token as we will further fine-tune the models. At last, we concatenate the two embedding vectors for the post title and description and feed them into a fully-connected layer. 


\subsubsection{Content of Code Blocks}\label{subsubsec:codeblock}
From our qualitative study, we find that there exists a close correlation between the intentions and certain types of content that posts may have. Code blocks are the most common container users may employ to drop supplementary resources that are in different formats or forms besides code snippets. As code block frequently appears in technical posts, utilizing the content types of code block as a feature may improve the performance of intention detection for technical posts.

However, as indicated by our qualitative study, code blocks can be used in different ways to present information. Moreover, previous works~\citep{li2020tagdc, wang2015tagcombine} consider code snippets from posts in online communities to be of low quality. These factors undermine the effectiveness of directly leveraging the code block content to help represent QA posts. Therefore, we propose to use the content categories as a feature for intention detection. According to the findings from our qualitative study (in \ref{subsec:structure-study}), we consider the content categories that frequently appear in the code blocks of QA posts.

~\response{R3.7}{
To automatically detect the content categories (i.e., natural language, code, error message, config, command line, and others) of code blocks, we constructed a multinomial Naive Bayes classifier. Our approach utilizes regular expressions to tokenize texts from code blocks, distinguishing identifiers, operators, and brackets. We then employed TF-IDF (term frequency-inverse document frequency) to transform tokens into numerical arrays, representing each token frequency across the dataset. Based on the document-term matrix of textual data in code blocks, we trained the classifier.}

~\response{R3.7}{
We constructed a code block dataset for training and evaluating of the classifier by sampling and annotating code blocks from our data dump. The code block dataset has 10k samples in total, which are unevenly distributed in different classes (i.e., natural language, code, error message, config, and command line). We randomly splitted the dataset into an 80\% training set, a 10\% validation\accepted{\heng{in R2.12 you used validation: keep consistency}} set, and a 10\% test set. We adopted grid-search to tune the hyperparameter (i.e., Additive smoothing parameter) of the Multinomial Naive Bayes classifier using the evaluation set. We utilized SMOTE resampling~\citep{chawla2002smote} to address class imbalance during the training process. Utilizing the classifier's probability outputs across predefined content types, we assessed accuracy by considering classes with probabilities surpassing 0.5. A correct prediction was registered when one or more classes aligned with the ground truth. The classifier achieved an accuracy of 83.3\% on the test set.
}

~\response{R3.7}{
The probability outputs serve as a feature of posts for the model to detect intentions\accepted{ \heng{feature for what purpose}}. Notably, we concatenate all the texts in all code blocks if a post contains more than one code block. As detecting content in code blocks is not the main focus of this work, we do not go into details here. The implementation of feature extraction and the classifier is included in our replication package.
}



\subsubsection{Other features}
\citet{beyer2020kind} constructed QA posts intention classifiers, and their experiments showed that some textual features were beneficial for the recognition of certain intentions of posts. Thus, we incorporate the features (i.e., Word Count, Readability and Sentiment) identified in their study to improve our performance.

\subsubsection{Feature Fusion}
We concatenate the embeddings of the title and description and feed the feature vector with $768 \times 2$ dimensions to a fully connected layer. Other features are also concatenated and then merged with a fusion layer. Finally, an output layer (see \ref{sec:model-train-infer}) is followed and outputs the probabilities of intentions.

\subsection{Model Training and Inference}\label{sec:model-train-infer}
\subsubsection{Multi-label Loss Function}
As we formulate the intention detection task as a multi-label multi-class classification task, we use a Sigmoid function as our output layer and adopt the Binary Cross Entropy loss (BCE loss) for each output node, which is between the target and the predicted probabilities. Therefore, the loss function for our model is the summation of the BCE losses of all output nodes over a batch of training data.


\subsubsection{Training \& Fine-Tuning}
In our experiments, we adopt two different training settings according to the research questions we proposed. The first setting is to freeze the parameter of the PTM and train a classifier based on the embeddings and other features of the posts. The second setting allows updates to the parameters of the pooler\accepted{\heng{pooling}\xingfang{not pooling.}} layer of the PTM and assigns different learning rates for different components of the framework. The details can be found in the next section.

\subsubsection{Cross-validation} \label{subsubsec:crossval}
\response{R2.12}{We used five-fold cross-validation to counter the limited size of the dataset, aiming to enhance the reliability of our evaluation. We randomly divided our annotated dataset into five folds. For each iteration, we used one fold as the test set and the other four folds as the training set. We randomly separated 1/8 of the training set as our validation set, which was used to calculate the loss for guiding the early stopping. In total, we have 784 samples. For each iteration, we used 157 samples as the test set, 502 as training data, and 125 as the validation set. This approach enabled us to assess models using all available data. Even though the test set of each interaction is relatively small, the test sets of the five iterations combined cover all the 784 instances in the dataset. Our final evaluation result is the aggregated performance over the combined test sets of the five iterations, increasing the reliability of our evaluation result.}

Each iteration had one fold of data as the test set and the other four folds as the training set, with a randomly selected 1/8 of the training set used for validation and to calculate the loss for early stopping. This allowed us to cover all 784 instances and improve the final evaluation's reliability.

\subsubsection{Prediction Refinement}
We map the output to categories with a threshold of $0.5$ when evaluating our model. However, we made some adjustments to the predicted labels to make them more reasonable. First, we find that there exist cases when the output probabilities of all classes are under the threshold. In such cases, we force the model to output at least one label by assigning the class with the highest probability as the detection result. 
Second, we eliminate any other predicted labels when the probability of the \textit{Others} class exceeds the threshold. This is because when the content covers multiple aspects of a programming topic or issue, the model may produce several output labels other than \textit{Others}.

\section{Evaluation} \label{sec:eval}

This section first introduces the dataset and metrics we use to evaluate our proposed intention detection framework. Organized along three research questions (RQs), we describe our experiments, aiming to have a better understanding of the characteristics of our proposed framework. Moreover, we analyze and summarize the results and findings.


\subsection{Evaluation Metrics}
As we formulate our intention detection task as a multi-class multi-label classification problem, we follow previous works on tag recommendation~\citep{zhou2017scalable, li2020tagdc, he2022ptm4tag} to use \textit{Precision@k, Recall@k, F1-score@k} to evaluate the performance of our approach. However, as our baseline models do not predict the probability for each class, we can not apply these metrics to them, which hinders the direct comparison with the baseline models. Therefore, we further employ the \textit{Micro F1 score} to get the overall performance over all classes, considering that posts of different categories take up different proportions of our dataset. 

\noindent \textbf{Precision@k, Recall@k, F1-score@k} evaluate the tag recommendation approaches on their performance predicting top-k tags. Our qualitative study
found that the posts usually have less than three intentions. Therefore, we set the value of k to 3. \textit{Precision@k} is the average ratio of the correctly predicted tags among the top-k labels. \textit{Recall@k} is the ratio of correctly predicted top-k tags among the ground truth tags.
As the value may be capped to be small, a modification is made to the equation  when the \textit{k} is smaller than the number of ground truth tags. And, \textit{F1-score@k} is the harmonic mean of \textit{Precision@k} and \textit{Recall@k}.

For each sample, the $Precision@k_i$ is defined by Equation~\ref{eq:eq1} and we average the value for all samples and get Equation~\ref{eq:eq2}. \accepted{\heng{all the equations for the metrics may not be needed as these are standard metrics.}}

\begin{small}
\begin{equation}\label{eq:eq1}
Precision@k_i = \frac{\left | Tag_i^{Pred} \cap Tag_i^{GT} \right |}{k}
\end{equation}

\begin{equation}\label{eq:eq2}
Precision@k = \frac{ {\textstyle \sum_{i=1}^{n} Precision@k_i} }{n}
\end{equation}
\end{small}

\textit{Recall@k} is defined by Equation~\ref{eq:eq3} and Equation~\ref{eq:eq4}.

\begin{equation}\label{eq:eq3}
Recall@k_i = \left\{\begin{matrix} 
  \frac{\left | Tag_i^{Pred} \cap Tag_i^{GT} \right |}{k} \;\;\;, \left | Tag_i^{GT} \right |>  k  \\  \\ 
  \frac{\left | Tag_i^{Pred} \cap Tag_i^{GT} \right |}{\left | Tag_i^{GT} \right | } \;\;\;, \left | Tag_i^{GT} \right |\le   k 
\end{matrix}\right. 
\end{equation}

\begin{equation}\label{eq:eq4}
Recall@k = \frac{ {\textstyle \sum_{i=1}^{n} Recall@k_i} }{n}
\end{equation}

, which is defined by: 

\begin{equation}\label{eq:eq5}
F1-score@k_i = \frac{2 \times Precision@k_i\times Recall@k_i}{Precision@k_i \ + Recall@k_i}
\end{equation}

\begin{equation}\label{eq:eq6}
F1-score@k = \frac{ {\textstyle \sum_{i=1}^{n} F1-score@k_i} }{n}
\end{equation}

\noindent \textbf{Micro Precision, Recall and F1-score} are commonly used to access the performance of a multi-class classifier when there exists more than one class and need to aggregate in some way. As our dataset is unbalanced, and the number of posts of different intention categories varies, we do not employ macro averaging for the scores. 
Micro average aggregation uses the normal version of scores by calculating with total numbers of True Positives (TP), True Negatives (TN), False Positives (FP), and False Negatives (FN) over all classes instead of for each class.

\noindent \textbf{Area Under Curve (AUC)} measures the degree of separability. It reflects the capability of a classifier to distinguish between classes. In our experiments, we adopt the one-vs-one configuration to compute the average AUC for all pairwise combinations of classes~\citep{hand2001simple}.

\noindent \textbf{Top-K accuracy} takes the k predictions with the highest probability to calculate the accuracy. It measures how likely the true label appears in the top-k predictions.

\subsection{Research Questions}

We propose the following three research questions (RQs) to assess the performance and understand the characteristics of the proposed framework.

\noindent \textbf{RQ1: Which pre-trained model (PTM) works best in our framework?}

\noindent \textbf{Motivation}
The transformer-based PTMs have been achieving promising results on various natural language~\citep{wolf2019huggingface} and computer vision tasks~\citep{dosovitskiy2020image, carion2020end}. The BERT architecture~\citep{devlin2018bert} gains great popularity due to its ability to achieve outstanding performance on various natural language processing tasks when trained with massive data. Prior work has applied different variants of BERT in software engineering tasks, such as tag recommendations for Stack Overflow posts~\citep{he2022ptm4tag}\accepted{\heng{citation}}.

However, the efficacy of using different PTMs varies according to the specific tasks and data according to previous works~\citep{von2022validity, yang2022aspect}. And according to evaluations from previous works~\citep{yang2022aspect}, domain-specific PTMs do not necessarily have better performances on domain-specific tasks. We do not have general guidance on what specific PTMs should be used in our circumstances. 
Therefore, in RQ1, we aim to comparatively evaluate the performance of our intention detection framework with the different PTMs.

\accepted{\heng{Move the following paragraphs to approach: We compare the performance of six variants of BERT in our framework... followed by the text below}}

\noindent \textbf{Approach}
In this RQ, We compare the performance of six variants of our intention detection framework with transformer-based PTMs.

Basically, the BERT architecture contains an encoder stack of transformer blocks. The original BERT is released in two sizes. We use the $\bf BERT_{base}$ in the experiment. The BERT$_{base}$ has 12 layers of transformer block with a hidden unit size of 768 and 12 self-attention heads in the encoder stack. In total, it contains 110M parameters and is trained with a large corpus of English data. In the following, we will be using BERT to denote the BERT$_{base}$ model.

Most other transformer-based PTMs inherit BERT architecture while adopting different training settings (e.g., tasks, hyper-parameters, data, etc.) to train according to their specific application scenarios. 
\textbf{RoBERTa}~\citep{liu2019roberta} modified some hyper-parameters and training tasks while maintaining the original BERT architecture. \textbf{ALBERT}~\citep{lan2019albert} further improve the original BERT by adopting parameter reduction techniques. \textbf{DistilBERT}~\citep{sanh2019distilbert} is a distilled version, which has 40\% fewer parameters while maintaining over 95\% of the BERT model. To process domain-specific texts in software engineering, which contain technical jargon that can not be properly processed by general language models,
\textbf{BERTOverflow}~\citep{tabassum2020code} is proposed with a named entity recognition technique. It is trained with sentences from Stack Overflow and can achieve better performance on domain-specific tasks. Further, there are also PTMs targeting software engineering tasks. Pre-trained with both natural language corpus and programming language data, \textbf{CodeBERT}~\citep{feng2020codebert} is able to generate embeddings for both forms of input data. It has been achieving promising results on several software-related downstream tasks~\citep{huang2021cosqa, mashhadi2021applying}.

We compare the performances of six variants of our framework with the PTMs mentioned above. We leverage the PTMs released in the online community Hugging Face~\citep{wolf2019huggingface} in our experiments. We use the pooler output of the PTMs, which corresponds to the representation of the first token. Since DistilBert is not pre-trained with the next sentence prediction task, there is no pooler output layer. Instead, we use the output of a linear classification head. During the training process, we fixed all parameters of the PTMs, and only updated the parameters of layers on top of the PTMs.

\noindent \textbf{Results and Analysis}
Table~\ref{tab:ptm_metrics_micro} shows the results of our experiments on different variants of our proposed intention detection framework with different PTMs. In terms of the Micro F1-score, the variants with BERT and RoBERTa models achieved the same performance (F1-score of 0.522) and outperformed other variants.
It may be worth noting that, \textbf{as a multi-class multi-label problem with seven classes, it is generally significantly more difficult to make accurate classifications than binary classification scenarios~\citep{sahare2012review}.}
The worst performance was achieved by the variant with BERTOverflow, a domain-specific PTM trained with Stack Overflow data, with a Micro F1-score of 0.340. The result seems counter-intuitive as its pre-training data is most relevant to our task and dataset. However, previous works~\citep{yang2022aspect, he2022ptm4tag} also found that this domain-specific PTM may perform worse than other general-purpose counterparts. The possible explanation may be that general-purpose PTMs were usually trained with larger data and have a better generalization ability. Compared with BERTOverflow, another domain-specific PTM (i.e., CodeBERT) achieved a moderate result, only inferior to the best ones by 5.7\%. This is likely due to the similarity between our input data and the training data of CodeBERT, as our input texts are sometimes a mixture of natural language and truncated codes\accepted{\heng{we did not include code in the input of BERT?}}, while CodeBERT is trained with both natural and programming languages. We also observed a performance loss when PTMs with fewer parameters were used: the distilled version of BERT (i.e., DistilBERT) compared with BERT. The variant with ALBERT was outperformed by the best two by 26.4\% in terms of Micro F1-score.

The results with F1-score@k scores further confirm the different performances. From Table~\ref{tab:ptm_metrics_atk}, we find that the variant with RoBERTa consistently outperformed others in terms of all F1-score@k, which indicates the predictions of this variant are of higher quality. However, the difference is insignificant between the two best variants: the differences of F1-score@k are 0.013, 0.048 and 0.033 when k is 1 to 3, respectively. The AUC score and Top-k accuracy show a similar trend as the F1-score, so we do not include them in the table.

\begin{table}[t]
    \centering
    \caption[AtKScore]{Comparison of variants of our framework with Micro F1-score. The highest scores and best variants are shown in \textbf{bold}.}
    \label{tab:ptm_metrics_micro}
    \resizebox{0.6\linewidth}{!}{
    \begin{tabular}{|c|c|c|c|} 
\hline
\multirow{2}{*}{Variant~} & \multicolumn{3}{c|}{Micro averaging}              \\ 
\cline{2-4}
                          & Precision      & Recall         & F1-score        \\ 
\hline
BERT             & 0.571          & \textbf{0.480} & \textbf{0.522}  \\ 
\hline
\textbf{RoBERTa}          & \textbf{0.597} & 0.465          & \textbf{0.522}  \\ 
\hline
ALBERT                    & 0.465          & 0.371          & 0.413           \\ 
\hline
DistilBERT                & 0.576          & 0.454          & 0.508           \\ 
\hline
BERTOverflow              & 0.402          & 0.295          & 0.340           \\ 
\hline
CodeBERT                  & 0.567          & 0.435          & 0.492           \\
\hline
\end{tabular}
}
\end{table}

\begin{table}[t]
    \caption[AtKScore]{Comparison of variants of our framework with different PTMs with \textit{Precision@k}, \textit{Recall@k} and \textit{F1-score@k}. The highest scores and best variants are shown in \textbf{bold}.}
    \label{tab:ptm_metrics_atk}
    \resizebox{\linewidth}{!}{
    \begin{tabular}{|c|c|c|c|c|c|c|c|c|c|} 
    \hline
    \multirow{2}{*}{Variant} & \multicolumn{3}{c|}{Precision} & \multicolumn{3}{c|}{Recall} & \multicolumn{3}{c|}{F1-score}  \\ 
    \cline{2-10}
                         & @1    & @2    & @3             & @1    & @2    & @3          & @1    & @2    & @3       \\ 
    \hline
    $BERT$          & 0.575 & 0.448 & 0.363          & 0.575 & 0.673 & 0.802       & 0.575 & 0.523 & 0.485    \\ 
    \hline
    \textbf{RoBERTa}              & \textbf{0.588} & \textbf{0.486} & \textbf{0.385}          & \textbf{0.588} & \textbf{0.740}  & \textbf{0.864}       & \textbf{0.588} & \textbf{0.571} & \textbf{0.518}    \\ 
    \hline
    ALBERT               & 0.459 & 0.383 & 0.332          & 0.459 & 0.576 & 0.736       & 0.459 & 0.447 & 0.445    \\ 
    \hline
    DistilBERT           & 0.570 & 0.465 & 0.375          & 0.570 & 0.706 & 0.836       & 0.570 & 0.545 & 0.503    \\ 
    \hline
    BERTOverflow         & 0.397 & 0.354 & 0.320          & 0.397 & 0.546 & 0.724       & 0.397 & 0.418 & 0.432    \\ 
    \hline
    CodeBERT             & 0.561 & 0.452 & 0.370          & 0.561 & 0.688 & 0.831       & 0.561 & 0.530 & 0.498    \\
    \hline
    \end{tabular}
    }
\end{table}


\begin{answer*}{to RQ1}{}
Our intention detection framework achieves the best performance with
the BERT variants 
\textbf{RoBERTa} and \textbf{BERT}. Generally, general-purpose PTMs work better than domain-specific counterparts in our intention detection framework, as they may be trained with larger data. PTMs with fewer parameters may suffer a performance loss.
\end{answer*}

\noindent \textbf{RQ2: Can our framework benefit from fine-tuning the PTMs? Compared with the baseline models, how effective is our intention detection framework?}

\noindent \textbf{Motivation}
In this research question, we have two goals: The first objective is to examine if the performance of our approach can be further improved by fine-tuning the PTMs with the intention detection task. \response{R1.8}{We chose two baselines for our study. The first one, proposed by~\citet{beyer2020kind}, uses a set of random forest binary classifiers for QA post intention detection. The second baseline is a convolution neural network (CNN)-based approach from~\citet{huang2018automating} which is designed for extracting intentions from GitHub issue reports.}

 
\noindent \textbf{Approach}
In RQ1, we fixed the parameters of PTMs 
and only updated the layers on top of them in backpropagation to examine the effectiveness of various PTMs. To answer this research question, we further fine-tune the pooler layer in the two best-performing PTMs (e.g., BERT and RoBERTa).
We then compare the fine-tuned models with our baseline approach. As the taxonomy for intentions from the baseline approaches differs from that of this work, we cannot directly compare the classification results. Therefore, we follow the implementations from the previous work and evaluate the approaches on our annotated dataset.~\response{R2.13}{We evaluate the baseline approaches with the same cross-validation process mentioned in Section~\ref{subsubsec:crossval}.}


\noindent \textbf{\textit{Baseline 1: Random Forest Binary Classifiers}} are used in~\citet{beyer2020kind} to classify Stack Overflow posts into seven intention categories. In their work, a set of features extracted from the posts serves as input to the machine-learning-based classifiers. The feature combinations mainly include \textit{N-gram} of the text or the part-of-speech tags (POS) of the text, word count, code snippets, and some other textual features (e.g., readability, sentiment). The authors conducted experiments on all feature combinations with a set of random forest binary classifiers and determined the best configurations for the task. This approach, to our knowledge, is the state-of-the-art work that proposed an automated approach for detecting QA post intentions. In our work, we follow the preprocessing and configurations from their work and train a set of random forest classifiers for our intention categories with our dataset as the baseline model.~\response{R2.13}{In our approach, individual random forest classifiers are dedicated to distinct intention categories, functioning as binary classifiers. These classifiers generate predictions specific to their assigned categories. To craft the multi-class multi-label classification output for each post, we merge the outputs from these classifiers. The final prediction for a post's intention categories is determined by aggregating the individual classifier's outputs.}


\response{R1.8}{\noindent \textbf{\textit{Baseline 2: A CNN-based approach}} is introduced by~\citet{huang2018automating} for the task of extracting intentions from issue reports on GitHub. This approach applies a CNN-based network and classifies sentences from issue reports into seven pre-defined intention categories. Batch normalization is integrated with the CNN layer to enhance training speed. In our work, we utilize the same CNN architecture while we substitute the cross entropy loss with the BCE loss to adapt to our task, which requires a multi-label output. We concatenate the title and description of posts and use the pre-trained GloVe word embeddings~\citep{pennington2014glove} to transform words into the corresponding vector representations as the input to the CNN model.
}

\noindent \textbf{Results and Analysis}
To better evaluate the performance of our proposed approach, we performed 10-fold cross-validation and calculated the metrics over all the posts in our dataset. Table \ref{tab:finetuned} shows the performances of the baseline models and two best-performing variants from RQ1 after fine-tuning. From the tables, we observe an overall improvement in performance: compared with the models without fine-tuning PTMs, the Micro F1-scores increase by 5.2\%, and 12.8\%, respectively. From Table~\ref{tab:tuned_auc}, we can observe the variations of the Precision@1, recall@1, and F1-score@1, which follow the same trend as the previous metrics. For these two variants after fine-tuning, the Top 1-3 accuracy ranges from 58.7\% to 84.8\% and 62.6\% to 87.8\%, respectively.

Since the pooler layer in PTMs is often used by downstream tasks in the pre-training stage, its parameters could be highly pertinent to the particular tasks, which undermines the quality of the output embeddings. This may explain the improvement in performance observed in our experiments. As 
by fine-tuning this layer with our task, the quality of embedding may be improved for our downstream task.
The performance improvement is reflected in the average AUC, with an increment of 3.0\% and 6.8\% for the two variants. 


\accepted{\heng{Discuss why the fine-tuning improves the performance (more specific context etc.)}}

\begin{table}
\centering
\caption{\response{R1.8: Table updated.}{}The performance of our approach after fine-tuning the PTMs compared with baselines. The values in parentheses indicate the absolute differences compared with results in RQ1.}
\label{tab:finetuned}
\resizebox{0.80\linewidth}{!}{
\begin{tabular}{|c|c|c|c|c|} 
\hline
\multirow{2}{*}{Model}                                                & \multicolumn{3}{c|}{Micro averaging}                                                                                                                                                                                              & \multirow{2}{*}{\begin{tabular}[c]{@{}c@{}}Average\\AUC~\end{tabular}}  \\ 
\cline{2-4}
                                                                      & Precision                                                                 & Recall                                                                    & F1-score                                                                  &                                                                         \\ 
\hline
BERT                                                                  & \begin{tabular}[c]{@{}c@{}}0.562\\(0.052$\uparrow$)\end{tabular}          & \begin{tabular}[c]{@{}c@{}}0.536\\(0.056$\uparrow$)\end{tabular}          & \begin{tabular}[c]{@{}c@{}}0.549\\(0.027$\uparrow$)\end{tabular}          & 0.754                                                                   \\ 
\hline
\textbf{RoBERTa}                                                      & \begin{tabular}[c]{@{}c@{}}\textbf{0.601}\\(0.004$\uparrow$)\end{tabular} & \begin{tabular}[c]{@{}c@{}}\textbf{0.577}\\(0.112$\uparrow$)\end{tabular} & \begin{tabular}[c]{@{}c@{}}\textbf{0.589}\\(0.067$\uparrow$)\end{tabular} & \textbf{0.787}                                                          \\ 
\hline
\begin{tabular}[c]{@{}c@{}}Baseline 1\\(Random Forest)\end{tabular} & 0.597                                                                     & 0.462                                                                     & 0.521                                                                     & 0.745                                                                   \\ 
\hline
\begin{tabular}[c]{@{}c@{}}Baseline 2\\(CNN-based)\end{tabular}     & 0.558                                                                     & 0.577                                                                     & 0.567                                                                     & 0.765                                                                   \\
\hline
\end{tabular}
}
\end{table}

\begin{table}
\centering
\caption{The performance after fine-tuning.}
\label{tab:tuned_auc}
\begin{tabular}{|c|c|c|c|} 
\hline
\multirow{2}{*}{Variant} & \multicolumn{3}{c|}{@1}                                                                                                                                                                                 \\ 
\cline{2-4}
                         & Precision                                                        & Recall                                                           & F1-score                                                          \\ 
\hline
BERT            & \begin{tabular}[c]{@{}c@{}}0.587\\(0.012$\uparrow$)\end{tabular} & \begin{tabular}[c]{@{}c@{}}0.587\\(0.012$\uparrow$)\end{tabular} & \begin{tabular}[c]{@{}c@{}}0.587\\(0.012$\uparrow$)\end{tabular}  \\ 
\hline
RoBERTa                  & \begin{tabular}[c]{@{}c@{}}0.626\\(0.038$\uparrow$)\end{tabular} & \begin{tabular}[c]{@{}c@{}}0.626\\(0.038$\uparrow$)\end{tabular} & \begin{tabular}[c]{@{}c@{}}0.626\\(0.038$\uparrow$)\end{tabular}  \\
\hline
\end{tabular}
\end{table}

\begin{answer*}{to RQ2}{}
By fine-tuning the pooler layer of PTMs with our annotated dataset, the performance of our intention detection framework is further improved, achieving a Micro F1-score of 0.589, Top 1-3 accuracy of 62.6\% to 87.8\%, and an average AUC of 0.787.
Our proposed approach outperforms the baselines. 
\end{answer*}

\noindent \textbf{RQ3: Can the content category of code blocks really help the detection of post intentions?}

\noindent \textbf{Motivation}
From our qualitative study, we found that the content of code block has a close correlation with the intention of posts. Thus, we implement a code block content classifier and employ the predicted probabilities of content categories as a feature for intention detection. In this research question, we examine the effectiveness of this feature and further validate the findings from our qualitative study from an experimental point of view.

\noindent \textbf{Approach}
We conduct an ablation study to investigate the importance of employing code block content as a feature. To answer this research question, we remove the code block classifier and modify our proposed framework to fit the dimensions of the input feature. We train and evaluate the ablated framework with 5-fold cross-validation to get an unbiased performance estimation.

\noindent \textbf{Results and Analysis}
Table~\ref{tab:ablation_micro} shows the results of our ablation study. When compared with the framework with the code block content classifier, we observed a minor performance loss of the ablated ones. However, the loss is not significant: The F1-scores and average AUC for BERT decrease by 1.1\% and 0.5\%\accepted{\heng{check if these values are updated}}. The F1-scores and average AUC for RoBERTa decrease by 2.3\% and 1.8\%\accepted{\heng{check these values too}}. The results confirm our assumption that the occurrence information of code block content can help the intention detection. However, the benefits of using this as a feature for intention detection may be undermined by the strong representation ability of efficient PTMs.

\begin{table}[!ht]
\centering
\caption{The results of ablation study.}
\label{tab:ablation_micro}
\begin{tabular}{|c|c|c|c|c|} 
\hline
\multirow{2}{*}{Ablated Model} & \multicolumn{3}{c|}{Micro averaging}                                                                                                                                                                         & \multirow{2}{*}{\begin{tabular}[c]{@{}c@{}}Average\\AUC\end{tabular}}  \\ 
\cline{2-4}
                               & Precision                                                          & Recall                                                             & F1-score                                                           &                                                                        \\ 
\hline
BERT                  & \begin{tabular}[c]{@{}c@{}}0.554\\(0.008$\downarrow$)\end{tabular} & \begin{tabular}[c]{@{}c@{}}0.533\\(0.003$\downarrow$)\end{tabular} & \begin{tabular}[c]{@{}c@{}}0.543\\(0.006$\downarrow$)\end{tabular} & \begin{tabular}[c]{@{}c@{}}0.753\\(0.001$\downarrow$)\end{tabular}     \\ 
\hline
RoBERTa                        & \begin{tabular}[c]{@{}c@{}}0.590\\(0.011$\downarrow$)\end{tabular} & \begin{tabular}[c]{@{}c@{}}0.560\\(0.017$\downarrow$)\end{tabular} & \begin{tabular}[c]{@{}c@{}}0.575\\(0.014$\downarrow$)\end{tabular} & \begin{tabular}[c]{@{}c@{}}0.773\\(0.014$\downarrow$)\end{tabular}     \\
\hline
\end{tabular}
\end{table}

\begin{answer*}{to RQ3}{}
The category of code block content can serve as a feature to help the detection of post intentions in our framework. However, the effectiveness is limited.
\end{answer*}

\section{Lessons Learned}\label{sec:industrial}

\subsection{Insights from Collaborative Industry Endeavors}\label{subsec:industrial}
Our research has been driven and conducted in close collaboration with an industry partner that specializes in gathering, analyzing, and recommending information from online technical communities. The industry partner's use cases, feedback, and involvement in the co-construction approach have provided valuable insights and contributions for building the taxonomy, designing and improving the intention detection tool, and adopting the outcomes in the industry environment.

\noindent \textbf{Driven by the industrial use cases} Our collaboration with our industrial partner enables us to have access to certain industrial use cases that motivate our research. Primarily, our partner is keen on enhancing content recommendations for their platform's end users. This endeavour involves correlating potential posts with user profiles crafted from users' info and their activity histories, which may reflect their varying levels of expertise. However, the existing post tags available focusing on technical topics fail to capture the intentions of posts that may be related to the level of expertise of users. For example, novice programmers are more interested in finding learning resources—a prevalent focus commonly often found in posts categorized under the \textit{Learning} intention. Conversely, this intention can serve as a proactive strategy to minimize the delivery of undesired content to specific user groups. Experienced users often perceive \textit{Learning} posts as repetitive or less engaging due to their advanced knowledge. Leveraging our intention detection approach, we aim to refine the recommendation system to address the distinct needs of both novices seeking learning materials and experienced users with more advanced requirements. This strategy enables a more engaging and enriching user experience, ensuring that content recommendations cater to diverse proficiency levels across the platform.

\noindent \textbf{Improving the generalizability}
The partnership has been helpful in identifying certain limitations in the existing intention taxonomies (e.g.,~\cite{allamanis2013and, beyer2020kind, treude2011programmers, beyer2014manual}\accepted{\heng{cite}}). Feedback from our partner indicates that certain categories may hold little significance due to a small number of related posts, as well as a lack of use cases and low generalizability.\accepted{\heng{not sure if ``lack of use case'' and ``low generalizability'' are at the same level of ``low significance'' or ``small portion of related posts''. Reorganize this sentence to make it clearer}} For instance, the category \textit{API-related}, commonly found in previous works on specific domains such as Android~~\citep{beyer2014manual}, may not be applicable to general domains that do not always involve API usage. Its relevance can also overlap with almost all other intentions, making it less generalizable. Therefore, it has been combined with a more general \textit{How-to} intention.

\noindent \textbf{Enhancing the practicability\accepted{\heng{Improving the usability of the taxonomy and tool?}}\accepted{\heng{The key is to highlight how the industry involvement contributes to the outcomes of the research}}}
In order to better serve the needs of our industrial partner, we have designed our taxonomy to improve the usability in an industry environment\accepted{\heng{avoid saying specific use cases, as it seems it lacks generalizability. Instead, say ``to improve the usability of the taxonomy in an industry environment''}}. For example, our partner has observed that beginner programmers often post elementary\accepted{\heng{elementary?}} programming questions, which experienced users may find repetitive. To address this, we have included a \textit{Learning} category in our intention taxonomy to categorize these types of posts, despite the fact that such posts are under-represented\accepted{\heng{despite the fact that such posts are under-represented}}. 
Furthermore, by combining intention categories with technical topics, we can direct questions to the appropriate domain experts more efficiently. For instance, identifying \textit{How-to} posts and pairing them with technical tags can help the recommendation system suggest related questions to domain experts who are willing to answer questions. 
While our intention taxonomy covers many use cases, there may be posts that don't fit into any of our categories. These posts are classified under the \textit{Other} category as our partner has not found any practical use or benefit for them in the recommendation system.


\noindent \textbf{Improving the implementation \accepted{\heng{Improving the implementation?}}}As mentioned previously, employees from our industrial partner have been involved in our intention annotation process. The involvement of these domain experts and developers makes our annotation results to be more accurate, trustworthy, and applicable. Furthermore, their input is also incorporated into the design, construction, and evaluation of our intention classification models. By assimilating their suggestions and opinions, we ensure that our models better align with the practical requirements of the industry, making them more relevant and applicable. 

\noindent \textbf{Continuous improvement with industrial adoption} The performance of our prototype model may be limited due to the shortage of well-annotated data, which requires significant manpower. However, our industry partner is integrating our taxonomy and classification approach into their platform to help their clients find relevant technical forum posts. We plan to enhance our model's performance by gathering more annotated data through end-user feedback. By doing so, we stand a good chance of improving the model's accuracy and effectiveness in a continuous and iterative manner.

\response{R1.5}{\subsection{Using pre-trained language models on forum post data}\label{subsec:takeaways_pretrained}} 
\response{R1.5}{\noindent \textbf{Pre-trained language models (PTMs) are effective in representing technical forum data.} In our study, we explored the efficacy of PTMs in representing technical forum data, specifically targeting the task of intention classification for technical posts. Our analysis revealed that PTMs, leveraging their advanced language understanding capabilities, excel at effectively capturing and representing the nuanced information within technical forum discussions. \mresponse{R2}{By utilizing PTMs, our proposed model demonstrated competitive performance in detecting post intentions, even without fine-tuning, when compared to baseline methods employing traditional feature extraction or basic word embedding.} This underscores the utility and proficiency of PTMs in handling textual data related to software. \mresponse{R2}{Therefore, we suggest that researchers and practitioners explore the utilization of PTMs across a spectrum of challenges within the realm of software engineering (e.g., uncover known issues in users' feedback for software maintenance).}} 

\response{R1.5}{\noindent \textbf{Fine-tuning PTMs can be expensive.} Our experimental results highlight the importance of fine-tuning the PTMs to achieve superior results for target downstream tasks. However, acquiring well-annotated data for fine-tuning PTMs for the targeted downstream task may be a resource-intensive endeavour. In our experiment, we had only 784 annotated posts available for fine-tuning and model evaluation. Consequently, to adapt the model to our tasks, we chose to exclusively update the pooler layer within the PTMs. When fine-tuning the PTMs containing an extensive number of parameters, data deficiency can potentially lead to overfitting during the fine-tuning process. In this situation, practitioners and researchers may consider selectively updating specific layers of the PTMs. This approach allows the model to update only a subset of the parameters to adapt to the downstream tasks, without jeopardizing the generalizability of the generated embeddings.}

\response{R1.5}{\noindent \textbf{Domain-specific PTMs may not perform better.} In our experiment, we compare variants of our proposed intention detection approach with both domain-specific and general-purpose PTMs. Contrary to the intuition that domain-specific PTMs, trained with software engineering (SE)-related data, would outperform their general-purpose counterparts in our task, our experimental results present a contrary outcome. Across various evaluation metrics, the general-purpose PTMs generally demonstrated superior performance over their domain-specific counterparts. This unexpected result prompts a reconsideration of the intuition that domain-specific PTMs inherently lead to better outcomes for tasks within a particular domain. The performance of PTMs may be jointly influenced by multiple factors, such as model complexity, pre-training corpus volume, etc. We encourage practitioners and researchers to evaluate both types of PTMs for their specific downstream tasks to attain optimal results.
}

\response{R1.5 R1.6? R3.4}{
\subsection{Recommendations for technical forum developers and contributors}\label{subsec:postprac}
}

\response{R1.5 R1.6? R3.4}{
\noindent \textbf{Using intention as a separate dimension to identify forum posts.} In addition to employing technical tags that categorize posts based on subject matter or technical topics, exploring the intentions behind forum posts presents an exciting chance to improve user experience and content relevance in technical online communities. Incorporating an intention dimension in the content organization of technical forums offers an added context, revealing the motivations behind users' inquiries. Consequently, our recommendation is for forum developers to integrate a dedicated tagging or categorization system focused on discerning the intentions behind individual posts. This integration would empower forum users to contribute to and discover posts that align with particular intentions, cultivating a more purposeful interaction within these online communities.
}

\response{R1.5 R1.6? R3.4}{
\noindent \textbf{Making good use of the code blocks.} In our manual study, we observed numerous instances of code snippet misplacements and misuse of code blocks within technical forum posts (e.g., inline codes are often mixed with other descriptions\accepted{\heng{give a brief example}}). The mixture of code snippets and pieces of natural language can pose significant challenges for existing recommendation systems or technical forum data analysis methods (e.g., tag recommendation, intention mining), which are mainly based on extracting features from the texts, to generate accurate results. Hence, we encourage technical forum contributors\accepted{\heng{contributors?}} to adhere to posting conventions, placing code snippets in code blocks, and marking inline code appropriately. This adherence will enable a more accurate presentation of their posts.
}

\response{R1.5 R1.6? R3.4}{
\noindent \textbf{Setting clear objectives before posting.} The intention taxonomy can serve as a thinking aid for individuals formulating their questions in technical forums. By leveraging this taxonomy, questioners can better structure their questions, leading to a clearer and more precise expression of their objectives. By remaining mindful of the question's intended purpose, questioners improve their ability to articulate issues effectively, benefiting both repliers and readers. A clearer delineation of objectives aids not only those providing answers but also the broader audience in comprehensively understanding the issue, enabling targeted and more helpful assistance. \mresponse{R2}{Therefore, we suggest that contributors to technical forums consider adopting this approach, as it has the potential to improve the efficiency and productivity of information exchange within the community.}
}

\section{Threats to Validity} \label{sec:threats}

\noindent \textbf{Internal Validity.} 
Manual study and annotation may be subject to the subjectivity and even bias of the authors. To \response{R2.16}{reduce} this bias, the two authors examine the data independently. In most cases, the agreement can be made. In case of disagreement, two authors discuss, and one other author is involved in helping reach a consensus. The involvement of other experts other than authors can further mitigate the threat.

\noindent \textbf{External Validity.} \response{R2.14}{The datasets used (i.e., post intention and code block dataset) are restricted to limited numbers of technical forums. There are many forums or communities in the domain of software and hardware systems. However, the datasets were extracted from a working system from our industrial partner. The data has good coverage of mainstream technical developer communities. Future works can validate the generalizability of our findings and our approach with new forum posts from different sources.}

\noindent \textbf{Construct Validity.}
\response{R1.9}{We used random sampling to split our dataset into folds for cross-validation, potentially causing overrepresentation or underrepresentation of certain data sources in the folds used for testing. These biased representations might influence the accuracy of our evaluation. However, we utilize all available data in our evaluation process, ensuring its robustness.}

\response{R1.9}{Regarding the utilization of PTMs, we employed the output from the first \textit{$<$CLS$>$} token to represent the post data. Various pooling strategies exist for PTMs, and choosing among them can significantly affect the performance of downstream tasks, thus influencing our assessment of PTMs. Nevertheless, the \textit{$<$CLS$>$} token encapsulates contextual information learned across the input sequence and commonly acts as an initial reference for downstream tasks. Our fine-tuning process, based on this token, further reduced this influence. Concerning the fine-tuning of PTMs, our initial attempt involved updating the parameters of the entire pre-trained models. However, due to the limited size of the dataset, we encountered challenges in achieving favorable training outcomes. Our subsequent approach focused solely on updating the pooler layer of the PTMs, which might not be the most optimal solution. Future works may explore additional fine-tuning strategies to enhance overall performance.}

~\response{R2.14}{In our approach to handling code blocks within post data, we developed a classifier specifically tailored for predicting content categories. The efficacy of this classifier may influence the accuracy of intention detection, potentially impacting the construct validity of our approach. Moreover, relying solely on content categories as a feature while disregarding the actual content might lead to the loss of information, considering that the text within code blocks may hold essential insights into the post's intention. Employing advanced feature extraction and representation techniques, such as leveraging text embedding techniques to process textual data and generate representations for code block contents, holds promise for achieving more precise intention detection results. However, the adopted classifier has lower computational costs than many embedding techniques, resulting in fewer resources for practical implementation in production environments, ensuring the scalability and feasibility of our approach. Therefore, the feature of code block content is incorporated into the framework intended to be implemented on the industry partner platform.}

\response{R1.9}{Additionally, we implemented the baseline approaches and assessed their performance using our dataset. Given that our tasks and taxonomy differ from those in the original studies, we adapted the approaches accordingly. However, this re-implementation introduces the potential for errors and bias in our research. To mitigate this, we referenced the source codes of the original implementations during the baseline approach implementation, striving to maintain consistency with the original work. This effort aims to enhance the construct validity of our study.}

\noindent \textbf{Conclusion Validity.}
\response{R1.9}{We labeled a limited number of posts, which may limit the reliability of our conclusion (the taxonomy of post intentions). To mitigate this, we sampled a statistically representative sample of 384 posts from our data dump to conduct our manual study. We further increased the number of posts for intention annotation to 784 posts, which is still limited. Using the small number of posts to train and evaluate our intention classification approach may threaten our conclusion about the performance of the model. To mitigate this issue, we use a 5-fold cross-validation for the model and calculate its performance by aggregating the results across the five test folds. Future work could involve expanding the annotated dataset and employing a larger testing set to evaluate the model.}

~\response{R3.8}{Moreover, the limited numbers of samples in certain intention categories (i.e., \textit{Review}, \textit{Learning}) impedes our ability to accurately evaluate and compare the performance of our models and baseline approaches in classifying these intentions, potentially affecting the robustness and reliability of our study's findings and conclusions regarding these specific intention categories, which presents a threat to the conclusion validity of our study. Initially, our dataset contained 384 samples. In an effort to expand its size to 784, our focus primarily centered on annotating more data while maintaining the original distribution of intention categories. \mresponse{R2}{We did not acquire additional annotated data to train our approach due to the costs of an extensive manual annotation process.} To enhance our models' performance and evaluation, future strategies might involve leveraging our approach to identify posts within these categories. Subsequently, a human verification process could be employed to augment the training set selectively, aiming to uphold the original data distribution while refining model accuracy over these less frequent intention categories.}

\section{Related Work} \label{sec:related}

\response{R1.8}{
\subsection{Mining Intentions from Developers' Discussions}
}
\label{subsec:relint}

The rapid growth of programming-related online communities has highlighted the need to better understand the characteristics and nature of online community posts. Therefore, more and more researchers have been focusing on mining and analyzing the content produced by software practitioners. Besides the technical aspects, intention can serve as an important factor in classifying and arranging technical posts in online communities. Many researchers have proposed different taxonomies for the post by manual analysis. Although some of the works did not explicitly mention the word \textit{intention}, we can tell that some of their categories are a description of the posting purposes. \citet{treude2011programmers} were the first ones to manually classify Stack Overflow posts into ten categories from an intention aspect. \citet{allamanis2013and} used topic modeling to analyze questions from Stack Overflow and found the correlation between question types and programming concepts and identifiers. Instead of studying all question types, \citet{beyer2014manual} focused only on the android development questions and summarized 8 question types. They further employed a k-NN classifier to classify questions. Similarly, \citet{rosen2016mobile} conducted a study on mobile application development posts and classified them into \textit{How, What, Why}. In~\citet{beyer2020kind}, researchers proposed a taxonomy based on previous taxonomies and tried to construct classifiers to automatically classify Android-related QA posts from Stack Overflow. \response{R1.8}{Besides the studies focusing on question-answering post data, researchers have conducted analyses to extract intentions from other software-related sources. For instance, \citet{huang2018automating} introduced a taxonomy of intentions specific to issue reports in GitHub projects. They developed a Convolutional Neural Network (CNN)-based approach to automatically categorize sentences into predefined intention categories.
\citet{lu2022miar} focused on app reviews and proposed a deep-learning-based framework to classify them into four intention categories.
}\textbf{In this work, we studied the characteristics of community posts (including a classification of post intentions) that cover multiple developer communities and consider inputs from the industry. In addition, we proposed an automated intention detection framework that outperforms the state-of-the-art baseline. }

\subsection{Tag recommendation for developer community posts}

Researchers have developed various approaches to fulfill the task of tag recommendation in the software engineering domain. These approaches can automatically propose tags (mostly in technical aspects) for software artifacts, software objects, etc.~\citep{wang2015tagcombine, wang2018entagrec++, al2010fuzzy}. Here, we only briefly introduce recent works on the tag recommendation for developer community posts or objects in software information sites. \citet{hong2017efficient} propose a tag recommendation method based on topic modeling. This method computes tag scores according to the document similarities and historical tag occurrence. \citet{liu2018fasttagrec} proposed FastTagRec, which is a neural network-based method that can infer tags for new postings accurately and fast. \citet{zhou2019deep} proposed four tag recommendation methods based on four contemporary deep learning approaches, among which TagCNN and TagRCNN work better than traditional approaches. TagDC~\citep{li2020tagdc} further improved the performance by leveraging deep learning techniques and collaborative filtering techniques.
\textbf{Our proposed intention detection framework can complement these tag recommendation approaches by providing a different perspective for locating relevant posts.}

\section{Conclusions} \label{sec:conclusions}

The ever-growing online developer communities demand more efficient and rational ways of organizing content and making recommendations for users. Our work is just under this background. In this work, we first conducted a qualitative study on a sampled dataset from an industrial source to understand the common posting practices in technical communities. We proposed an intention taxonomy of technical posts by seeking feedback from our industrial partner and referring to previous studies. With this taxonomy, we manually annotated posts and analyzed the correlation between post intention type and post content. 
Based on the findings from the qualitative study, we proposed an intention detection framework that utilizes transformer-based pre-trained language models. We further examine the characteristics of the framework with three research questions, from which we validated the effectiveness of our approach compared with a baseline model and confirmed the relevance of code block content and post intention can be utilized and thus boost the intention detection task.
Our taxonomy of post intentions and automated detection framework may be leveraged by technical forum maintainers or third-party tool developers to improve the organization and search of relevant posts on technical forums.
To expand on our findings, future research could involve creating a more extensive post-intention dataset, or assessing the impact of utilizing post intents for enhancing post searches or recommendations.



\section*{Acknowledgements}

We would like to gratefully acknowledge the Mitacs-Accelerate program and the Natural Sciences and Engineering Research Council of Canada (NSERC) for funding this project.

%
\section*{Conflicts of Interests}
The authors have no competing interests to declare relevant to this article's content.

\section*{Data Availability Statements (DAS)}
We have released our annotated dataset and code in the supplementary material package hosted on a GitHub repository\footnote{\label{supmat}Supplementary material package: \newline\url{https://github.com/mooselab/suppmaterial-TechnicalPostIntention}}.

\bibliographystyle{natbib}
\bibliography{main}   


\end{document}